\documentclass[english]{elsarticle}
\usepackage{lmodern}
\usepackage[OT1]{fontenc}
\usepackage[latin9]{inputenc}
\usepackage[a4paper]{geometry}
\usepackage{color}
\usepackage{array}
\usepackage{booktabs}
\usepackage{multirow}
\usepackage{amsmath}
\usepackage{amssymb}
\usepackage{graphicx}
\usepackage{setspace}
\makeatletter

\providecommand{\tabularnewline}{\\}

\usepackage{nicefrac}
\usepackage{hyperref}
\hypersetup{colorlinks=true,citecolor=blue}

\@ifundefined{showcaptionsetup}{}{%
 \PassOptionsToPackage{caption=false}{subfig}}
\usepackage{subfig}
\makeatother

\usepackage{babel}
\begin{document}

\begin{frontmatter}{}

\title{\textbf{Lattice Boltzmann Models for Micro-tomographic Pore-spaces}}

\author{Parthib Rao and Laura Schaefer}

\author{Rice University, Houston TX}
\begin{abstract}
The lattice Boltzmann method (LBM) is a popular numerical framework
to investigate single and multiphase flow though porous media. For
estimation of absolute permeability based on micro-tomographic images
of the porous medium, the single-relaxation time (SRT) collision model
is the most widely-used, although the multiple-relaxation-time (MRT)
collision model also has recently acquired wider usage, especially
for industrial applications. However, the SRT collision model and
a sub-optimal choice of the MRT collision parameters can both lead
to permeability predictions that depend on the relaxation time, $\tau$.
This parametric dependence is nonphysical for Stokes flow in porous
media and also leads to much larger number of iterations required
for convergence. In this paper, we performed a systematic numerical
evaluation of the different sets of relaxation parameters in the D3Q19-MRT
model for modeling Stokes flow in 3-D microtomographic pore-spaces
using the bounceback scheme. These sets of parameters are evaluated
from the point of view of accuracy, convergence rate, and an ability
to generate parameter-independent permeability solutions. Instead
of tuning all six independent relaxation rates that are available
in the MRT model, the sets that were analyzed have relaxation rates
that depend on one or two independent parameters, namely $\tau$ and
$\Lambda$. We tested elementary porous media at different image resolutions
and a random packing of spheres at relatively high resolution. We
observe that sets of certain specific relaxation parameters (Sets
B, D, or E as listed in Table \ref{tab:Different-sets-of}), and $\tau$
in the range $\tau\in[1.0,1.3]$ can result in best overall accuracy,
convergence rate, and parameter-independent permeability predictions.
\end{abstract}

\end{frontmatter}{}

\section{Introduction\label{sec:Introduction}}

Flow and transport in reservoir rocks and other porous media are strongly
dependent on the properties of the pore-space, such as the geometry,
distribution, and connectivity. Laboratory-based core analysis are
the main tools for studying and quantifying rock properties for practical
and scientific applications. Over the past decade, high-resolution,
three-dimensional (3-D), X-ray microtomography imaging and processing
capabilities have become widely available \cite{wildenschild2013x}.
These advances have enabled one to obtain a 3-D reconstructed pore-space
domain which serve as an input for the direct numerical simulation
of the underlying fluid and transport processes. In this approach,
also commonly called Digital Rock Physics (DRP) in the oil and gas
industry, transport properties of rocks, such as absolute (intrinsic)
permeability, relative permeability, dispersivity, formation factor,
etc., are computed on geometries constructed from 3-D microtomographic
images \cite{berg2017industrial}. In this study, we focus on the
permeability of a porous medium for single phase flow for Newtonian
fluids.

For direct numerical simulation of flow in microtomographic pores-spaces,
the lattice Boltzmann method (LBM) has emerged as a method of choice
\cite{Ferrol1995,fredrich2006predicting,khirevich2015coarse,manwart2002lattice,martys1996simulation,saxena2017references,ZARETSKIY20101508}.
After imaging, a cubic voxel in the 3-D image is segmented such that
each voxel is labeled as either a fluid or a solid. The resulting
fluid-solid interface, henceforth called the solid boundary, is inherently
rough and stair-step patterned, and which represents the surface roughness
of the medium at a length scale comparable to the resolution of the
image. Since the pore-space is already discretized via the voxels,
the segmented binary image lends itself immediately to numerical simulations
via the LBM without the additional step of pre-processing, such as
surface extraction and volume meshing that is needed in traditional
CFD. Pre-processing techniques, especially for tortuous porous geometries,
are tedious and therefore difficult to scale and automate. Besides
this advantage, there are other well-recognized benefits of the LBM
approach, such as the localized nature of the algorithm that enables
parallelization and good performance scalability, both essential features
for rapid and routine analysis of very large $(>1024^{3})$ images
with minimal user interface. One of the main challenges, however,
is the is the accuracy of the predictions.

The accuracy of the permeability predictions on pore-spaces obtained
from imaging depend strongly on the the sampling rate (minimum voxel
size) and the thresholding values used for the segmentation procedure
\cite{bultreys2016imaging}. For a segmented image of a given voxel
resolution, within the LBM framework, the overall accuracy of the
permeability predictions depends on boundary schemes for the solid
boundary, inlet and outlet boundary conditions, simulation parameters,
and techniques used to initiate the flow within the porous sample
(body-force driven or pressure gradient). Numerical artifacts arising
due to each the above factors contribute to a varying degree to the
overall inaccuracy.

The numerical technique used to enforce a zero (no-slip) velocity
condition on the solid boundary using the so-called bounceback (BB)
scheme is one of the largest sources of errors. Bounceback boundaries
are particularly simple to implement and have played a major role
in making LBM popular. The simplicity lies in that one simply needs
a particular voxel to be labeled as a solid (obstacle) and no additional
numerical procedure is required. Although robust and mass conserving,
a well-known defect of the bounceback scheme is the fact that, when
used conjunction with the single-relaxation-time (SRT) collision model,
the exact physical location of the solid boundary depends on the relaxation
time $\tau$, a LBM parameter that numerically sets the kinematic
viscosity, $\nu$ of the fluid via the relation $\nu=(\tau-0.5)\delta t/3$.
This spatial error is proportional to $(\tau-\delta t/2{}^{2})$ \cite{d2009viscosity,ginzbourg1994boundary,ginzburg2003multireflection}.
As an example, for Poiseuille flow in a straight channel, the theoretical
permeability value is $k=H^{2}/12,$ while the error in the permeability
obtained numerically with BB-SRT method is equal to $(4\nu^{2}-\delta t/12)$
\cite{ginzburg2003multireflection}. Because $\tau$ controls both
the kinematic viscosity and solution accuracy, the computed permeability
is said to be \emph{viscosity-dependent}. Physically this means that
the effective pore size is a function of viscosity, a clear violation
of physical behavior. In addition to the nonphysical nature of the
$k-\nu$ dependence, there are several consequences of this parametric
dependence of practical import.\textcolor{black}{{} Firstly, since the
non-dimensional solutions of Stokes flow do not depend on viscosity,
p}erforming simulations with BB-SRT scheme with different values of
$\tau$ will yield different permeability values of the porous media.
Secondly, since the LBM is not as efficient for steady flows, a practical
method to improve convergence towards the steady state is to use large
values of $\tau$ or $(\nu)$. \textcolor{black}{However,} for $\tau\gg1,$
the numerical errors increase quadratically and therefore not advisable
especially for porous media flow where since the surface to volume
ratio is large. Such \emph{parametric dependence} of permeability
leads to subjectivity or operator bias, much similar to the subjectivity
arising due to the thresholding values used in the image segmentation
step- a challenge recognized in the DRP community \cite{bultreys2016imaging,wildenschild2013x}. 

The first and most obvious method to improve to improve the prediction
accuracy and reduce $k-\nu$ dependence, is to increase the voxel
resolution. This approach is not always practical due to constraints
of the imaging equipment. The second method is to employ more accurate
wall-boundary schemes, such as the multi-reflection scheme, spatial-interpolation
based schemes, and other second-order accurate interpolation-based
techniques for curved boundaries \cite{ginzburg2003multireflection,bouzidi2001momentum,xu2016novel}.
Although these provide improvement in accuracy, these schemes require
that the location of the solid boundary surface be known in reference
to the underlying voxel grid/lattice \cite{ahrenholz2006lattice,fattahi2016lattice,pan2006evaluation,maier2010lattice}.
The location of the boundary surface is quantified by the so-called
sub-grid or wall distance, usually denoted as $q_{i}(\boldsymbol{x}),$
which represents the projected distance from the boundary voxel to
the solid boundary surface along the lattice links. Wall distances
can be evaluated relatively easily for analytic surfaces, such as
spheres, cylinders, and sphere packings, but can become tedious for
microtomographic images due to the additional pre-processing steps,
such as reconstructing a boundary surface using a smoothing (iso-surface
extraction) operation based on the marching-cubes algorithm \cite{maier2010lattice,young2008efficient,ZARETSKIY20101508}.
With the projected imminent increase in imaging capabilities, with
image datasets as large as $2000^{3}$ becoming available, the bounceback
scheme remains an utilitarian option for microtomographic image based
pore-spaces, especially when large numbers of samples have to analyzed
rapidly from various regions of the reservoir/wells. It should also
be noted that, with the SRT model, the permeability depends on the
viscosity even if more accurate boundary schemes are employed \cite{d2009viscosity}.

Therefore, given an \emph{a priori} choice of the bounceback scheme,
the third option to improve the accuracy and produce viscosity-independent
permeability is to use more complex collision models, such as the
multi-relaxation-time (MRT) \cite{d2002multiple,lallemand2000theory,luo2011numerics}
or two-relaxation-time (TRT) collision models \cite{d2009viscosity,ginzburg2006variably,ginzburg2008two}.
The essential idea of these collision models is that, unlike the SRT
model, the MRT/TRT models provide additional free relaxation parameters
- six in the case of MRT and two in case of TRT-that can be tuned
to obtain more accurate or stable solutions \cite{luo2011numerics}.
When the MRT/TRT collision models are applied with the bounceback
condition, they provide a mechanism to correct for the boundary error
(and obtain viscosity-independent results) by tuning the relaxation
parameters, the choice of which is essentially empirical for non-trivial
porous media, such as reservoir rocks. However, it should be noted
that the mere application of MRT/TRT model, instead of the SRT model,
may not result in viscosity-independent results unless the relaxation
parameters are carefully chosen. In fact, a sub-optimal choice of
the MRT relaxation parameters does not actually produce to viscosity-independent
results as can be seen several recent works \cite{eshghinejadfard2016calculation,narvaez2010quantitative,saxena2017references}.
Therefore, from an application-oriented user perspective, the question
remains: \emph{which set of MRT relaxation parameters would provide
the most accurate and viscosity-independent (or parameter-independent)
permeability prediction for a certain class of porous media/rocks?}

To address this concern, in this paper, we perform a systematic numerical
evaluation of the different sets of relaxation parameters within the
MRT framework for modeling Stokes flow in microtomographic pore-spaces.
These sets of relaxation parameters are evaluated from the point of
view of accuracy and an ability to generate viscosity-independent
solutions. For this purpose, we study two elementary porous media
(channels with circular and triangular cross sections) and a high-resolution,
random, dense packing of identical spheres, a first-order approximation
for sandstone rock. The rest of the paper is organized as follows.
In Sec. \ref{sec:The-lattice-Boltzmann}, we describe the essential
details of the LBM collision models, detail the various parameterization
choices for the MRT model, and briefly describe the boundary conditions
used in this study. In Sec. \ref{sec:Numerical-Tests}, we discuss
the result obtained from the numerical experimentation on various
porous geometries and in Sec. \ref{sec:Conclusions}, we present our
conclusions.

\section{The Lattice Boltzmann Equation\label{sec:The-lattice-Boltzmann}}

Compared to the classical approach of directly solving the Navier-Stokes
equation to model hydrodynamics, the lattice Boltzmann equation (LBE)
is based on kinetic theory and the Boltzmann equation \cite{guo2013lattice,kruger2017lattice,luo2010lattice}.
The LBE solves the discrete Boltzmann equation on a space $\boldsymbol{x}$
that is discretely represented as a lattice $\delta x\mathbb{Z}^{d}$
in \emph{d} dimensions with a lattice constant $\delta x$, and where
the time $t$ is discretized with a constant time step-size $\delta t$,
i.e., $t_{n}\in\delta t\mathbb{N}_{0}$ with $\mathbb{N}_{0}={0,1,...}$.
The continuous velocity-space of the Boltzmann equation is approximated
by a set of discrete velocities $\mathbb{V}\equiv\{\boldsymbol{c}_{i}|i=0,1,\cdots,b\}$
that it is symmetric, $\mathbb{V}=-\mathbb{V}$ and $c_{0}=\boldsymbol{0}$
\cite{luo2011numerics}. In the LBE, the phase-space $(\boldsymbol{x}-\boldsymbol{c})$
and the time discretization are coupled such that for any $\boldsymbol{c}_{i}\in\mathbb{V}$,
if $\boldsymbol{x}$ is a node, then $\boldsymbol{x}+\delta t\boldsymbol{c}_{i}$
is also a node of the lattice. Such a lattice is usually denoted as
D\emph{d}Q\emph{q,} where $d$ is number of space dimension and $q=b+1$
represents the total number of discrete velocities in the set. On
such a discrete system, the evolution of the distribution functions
$f_{i}(\boldsymbol{x})$ is given by:
\begin{equation}
f_{i}(\boldsymbol{x}+\boldsymbol{c}_{i}\delta t,t+\delta t)=f_{i}(\boldsymbol{x},t)+C_{i}\delta t+F_{i}\delta t\label{eq:LBE}
\end{equation}
where $f_{i},C_{i},$ and $F_{i}$ are \emph{q}-dimensional (column)
vectors. Eqn. \ref{eq:LBE} consists of three distinct terms: the
term on the left-hand side describes the free flight (streaming) of
the distributions from one computational (lattice) node to the other
according to their discrete velocities $\boldsymbol{c}_{i},$ $C_{i}$
describes the local changes to distributions due to inter-particle
collisions, and $F_{i}$ describes the changes (sources) of momentum
density due to external body forces. The most widely used lattice
structure for isothermal hydrodynamics is the D2Q9 lattice in 2-D
and D3Q19 and D3Q15 lattices in 3-D. In this work we use the D3Q19
lattice.

\subsection{Collision Models\label{subsec:Collision-Models}}

The most widely used collision model is the so-called Bhatnagar-Gross-Krook
(BGK) model, which assumes that all the distributions relax to a local
equilibrium value at a single rate characterized by a relaxation time
$\tau,$ 
\begin{equation}
\boldsymbol{C}^{BGK}\equiv-\frac{1}{\tau}\left[\boldsymbol{f}(\boldsymbol{x},t)-\boldsymbol{f}^{(eq)}(\boldsymbol{x},t)\right]\label{eq:collision}
\end{equation}
Since only one relaxation time is used, the BGK model is also termed
the single-relaxation-time (SRT) model. The local equilibrium is given
by the Maxwell-Boltzmann distribution, which is itself a function
of the local hydrodynamic variables, $f_{i}^{eq}=f_{i}^{eq}(\rho,\boldsymbol{u})$.
The functional form of the equilibrium distributions is obtained by
expanding the Maxwell-Boltzmann distribution up to second-order in
local flow velocity:

\begin{equation}
f_{i}^{(eq)}(\boldsymbol{x},t)=w_{i}\rho(\boldsymbol{x},t)\bigg(1+\frac{(\boldsymbol{c}_{i}\cdot\boldsymbol{u})}{c_{s}^{2}}+\frac{(\boldsymbol{c}_{i}\cdot\boldsymbol{u})^{2}}{2c_{s}^{4}}-\frac{(\boldsymbol{u}\cdot\boldsymbol{u})}{2c_{s}^{2}}\bigg)\label{eq:BGK-EDF}
\end{equation}
The relaxation time is related to the fluid's kinematic viscosity
via $\nu=c_{s}^{2}(\tau-0.5)\delta t=c_{s}^{2}(\frac{1}{s_{\nu}}-0.5)\delta t$,
where $s_{\nu}=\frac{1}{\tau}$ is the relaxation rate. Note that
the terms 'relaxation rate' and 'relaxation parameters' (in case of
MRT/TRT models) are used interchangeably.

\subsubsection{MRT Collision Model\label{subsec:MRT-Collision-Model}}

In contrast to the SRT model where all distributions relax at a single
rate, in the multiple-relaxation-time (MRT) model, the collision process
occurs in the moment space, where each moment of the distribution
can relax towards the local equilibrium at its own independent relaxation
rate. The MRT collision model is based on the generalized lattice
Boltzmann equation (GLBE) method, where while the streaming of the
populations occurs in the velocity-space, the collision occurs in
the moment-space, which is spanned by the orthogonal eigenvectors
basis of the collision operator. The MRT collision operator is expressed
as:

\begin{equation}
\boldsymbol{C}^{MRT}\equiv-\mathbf{M}^{-1}S\mathbf{M}[{\rm \boldsymbol{f}}-{\rm \boldsymbol{f}^{eq}}]=-\mathbf{M}^{-1}S[{\rm \boldsymbol{m}}-{\rm \boldsymbol{m}^{eq}}]\label{eq:MRT Collision}
\end{equation}
where $\mathbf{M}$ is a $q\times q$ matrix that linearly transforms
the $q$-dimensional vector of distribution function ${\rm \boldsymbol{f}},$
from the velocity-space to a $q$-dimensional vector ${\rm \boldsymbol{m}},$
in the moment-space: 

\begin{equation}
{\rm \boldsymbol{m}}=\boldsymbol{{\rm M}}\cdot{\rm \boldsymbol{f}}\quad\boldsymbol{{\rm f}}={\rm \boldsymbol{M}^{-1}}\cdot\boldsymbol{{\rm m}}\label{eq:momentTransformation}
\end{equation}
and the LBE-MRT equation is:
\begin{equation}
\boldsymbol{{\rm f}}(\boldsymbol{x}+\boldsymbol{c}_{i}\delta t,t+\delta t)={\rm \boldsymbol{f}}(\boldsymbol{x},t)-\mathbf{M}^{-1}S[{\rm \boldsymbol{m}}-{\rm \boldsymbol{m}^{eq}}]\delta t\label{eq:LBE-MRT}
\end{equation}
The rows of the transformation matrix $\mathbf{M},$ for the D3Q19
model consist of 19 orthogonal basis vectors, can be found in \cite{d2002multiple}.
The corresponding nineteen moments $\boldsymbol{m}$ of the distributions
are arranged in the following sequence:
\[
\mathbf{m}\equiv(\rho,e,\epsilon,j_{x},q_{x},j_{y},q_{y},j_{z},q_{z},3p_{xx},3\pi_{xx},p_{ww},\pi_{ww},p_{xy},p_{yz},p_{xz},m_{x},m_{y},m_{z}).
\]
Among the 19 moments, the $0$th-order moment corresponds to mass
density $m_{0}=\rho$ and the three 1st-order moments correspond to
momentum $m_{3,5,7}=j_{x,y,z}=\rho_{0}u_{x,y,z}.$ The six $2$nd-order
moments: $m_{1}=e,\,m_{9}=3p_{xx},\,m_{11}=p_{ww}=p_{yy}-p_{zz}$
and $m_{13,14,15}=p_{xy,yz,xz}$ correspond to part of the kinetic
energy, the diagonal, and off-diagonal elements of the viscous stress
tensor, respectively. Moments higher than second-order are called
as higher-order moments (also ghost moments) and they do not have
a clear physical connection to incompressible Navier-Stokes limit
of hydrodynamics, but are characterized by their dependence on the
gradient of the conserved moments. These higher-order moments are
the six 3rd-order moments $m_{4,6,8}=q_{x,y,z},\,m_{16,17,18}=m_{x,y,z}$
and three 4th-order moments $m_{2}=\epsilon,\,m_{10}=3\pi_{xx},\,m_{12}=\pi_{ww}$.
We refer to Table \ref{tab:D3Q19-Non-conserved-Moments} that summarizes
these relations.

The equilibrium moments are $m_{0}=m_{0}^{eq}=\rho=\sum_{i}f_{i}$
and $m_{3,5,7}=m_{3,5,7}^{eq}=j_{x,y,z}=\sum_{i}\boldsymbol{c}_{i}f_{i}$
since density and momentum are conserved during the collision process.
The local equilibrium moments for the non-conserved moments, which
are themselves functions of the conserved moments, are given as \cite{d2002multiple,pan2006evaluation}:

\begin{align}
m_{1}^{eq} & =m_{e}^{eq}=11\rho+\frac{19}{\rho_{0}}(j_{x}^{2}+j_{y}^{2}+j_{z}^{2})\nonumber \\
m_{2}^{eq} & =m_{\epsilon}^{eq}=3\rho-\frac{11}{2\rho_{0}}(j_{x}^{2}+j_{y}^{2}+j_{z}^{2})\nonumber \\
m_{4}^{eq}=m_{q_{x}}^{eq} & =-\frac{2}{3}j_{x},\quad m_{6}^{eq}=m_{q_{y}}^{eq}=-\frac{2}{3}j_{y},\quad m_{8}^{eq}=m_{q_{z}}^{eq}=-\frac{2}{3}j_{z}\nonumber \\
m_{9}^{eq}=3p_{xx}^{eq}=\frac{1}{\rho_{0}}(j_{x}^{2}-j_{y}^{2}-j_{z}^{2}), & m_{11}^{eq}=p_{ww}=\frac{1}{\rho_{0}}(j_{y}^{2}-j_{z}^{2}),\nonumber \\
m_{13}^{eq}=m_{p_{xy}}^{eq}=\frac{1}{\rho_{0}}j_{x}j_{y},\,\, & m_{14}^{eq}=m_{p_{yz}}^{eq}=\frac{1}{\rho_{0}}j_{y}j_{z},\,\,m_{15}^{eq}=m_{p_{xz}}^{eq}=\frac{1}{\rho_{0}}j_{x}j_{z},\label{eq:MRT Equi}\\
m_{10}^{eq}=m_{\pi_{xx}}^{eq}=-\frac{1}{2}p_{xx}^{eq}, & \,\,m_{12}^{eq}=m_{\pi_{ww}}^{eq}=-\frac{1}{2}p_{ww}^{eq},\nonumber \\
 & m_{16,17,18}^{eq}=m_{x,y,z}^{eq}=0\nonumber 
\end{align}
The relaxation rate matrix, $\mathbf{S}$, is a $q\times q$ diagonal
matrix which consists of the relaxation rate of each corresponding
moment, and which are the eigenvalues of the collision operator $\boldsymbol{\mathbf{M}^{-1}\mathbf{S}\mathbf{M}}$:

\begin{align}
\mathbf{S} & ={\rm diag}(0,s_{e},s_{\epsilon},0,s_{q},0,s_{q},0,s_{q},s_{\nu},s_{\pi},s_{\nu},s_{\pi},s_{\nu},s_{\nu},s_{\nu},s_{m},s_{m},s_{m})\label{eq:MRT params}
\end{align}
The relaxation rates $\{s_{i}\}$ determine the transport coefficients
in the system. The zero relaxation rates correspond to conserved moments
$\{\rho,j_{x},j_{y},j_{z}\}$ while the non-zero relaxation rates
are:
\[
s_{1}=s_{e},\,\,s_{2}=s_{\epsilon},s_{4}=s_{6}=s_{8}=s_{q},
\]
\[
s_{10}=s_{12}=s_{\pi},\,\,s_{9}=s_{11}=s_{13}=s_{14}=s_{15}=s_{\nu}
\]
\[
s_{16}=s_{17}=s_{18}=s_{m}
\]
The relaxation rate $s_{\nu}=\frac{1}{\tau}$ is related to the kinematic
viscosity $\nu$ via $\nu=c_{s}^{2}\big(\frac{1}{s_{\nu}}-0.5\big)\delta t$
where for athermal hydrodynamics $c_{s}^{2}=c^{2}/3$, where $c\equiv\delta x/\delta t$.
The remaining relaxation rates, $s_{e},s_{\epsilon},s_{q},s_{\pi},$
and $s_{m}$ can be set in the range $0<s_{i}<2$ in order to improve
the accuracy and stability. Simplifications can be made for the equilibrium
moments of the non-conserved quantities depending upon the flow at
hand. For example, for Stokes flow, the non-linear terms in velocity
can be ignored \cite{pan2006evaluation}. Another example is to set
equilibrium moments of the higher order moments $m_{2,4,6,8,16,17,18}$
to zero since these do not have physical link to the incompressible
Navier-Stokes behavior \cite{ahrenholz2006lattice}. In this work,
for the sake of generality, we use equilibrium moments as listed in
Eqn \ref{eq:MRT Equi}.

The above discussion was valid for flows initiated by pressure gradients,
or by shear, i.e. flows in the absence of external body forces. In
the following, we describe the methodology to include external body
forces, such as pressure gradient or gravity, within a MRT framework.
Given an external body force density $\boldsymbol{F}=\rho_{0}\boldsymbol{a}$,
where $\boldsymbol{a}$ is the acceleration due to applied force,
the effects of this force can be incorporated in LBE by either modifying
the equilibrium distribution function or by adding an explicit forcing
or a source term, $F_{i}$ \cite{guo2013lattice}. In this study,
we employ the method suggested by Guo et al., where LBE-MRT equation
with the force term is given as \cite{guo2007discrete,guo2013lattice}:

\begin{equation}
f_{i}(\boldsymbol{x}+\boldsymbol{c}_{i}\delta t,t+\delta t)=f_{i}(\boldsymbol{x},t)-M^{-1}\bigg(S[m_{i}-m_{i}^{eq}]-(I-\frac{1}{2}S)m_{i}^{F}\bigg)\label{eq:MRT-LBE}
\end{equation}
where $I$ is the $q\times q$ identity matrix and $\boldsymbol{m}^{F}=m_{i}^{F}=M\cdot F_{i}$
are the moments of the forcing term $F_{i}=\left\{ F_{0},F_{1},\cdots F_{b}\right\} ^{T}$
which is related to the Cartesian components of body force $\boldsymbol{F}=\{F_{x},F_{y},F_{z}\}$
as:

\begin{equation}
F_{i}=w_{i}\left(\frac{\boldsymbol{c}_{i}-\boldsymbol{u}^{eq}}{c_{s}^{2}}+\frac{\left(\boldsymbol{c}_{i}\cdot\boldsymbol{u}^{eq}\right)\boldsymbol{c}_{i}}{c_{s}^{4}}\right)\cdot\boldsymbol{F}\label{eq:Guo_Forcing}
\end{equation}
Note that the velocity that is used to calculate the above forcing
term and equilibrium moments in Eqn. \ref{eq:Guo_Forcing} is \emph{shifted},
i.e., $\rho\boldsymbol{u}^{eq}=\sum_{i}f_{i}\boldsymbol{c}_{i}+\frac{\rho_{o}\boldsymbol{a}\delta t}{2}$.
The velocity field $u^{eq}$ also satisfies the Navier-Stokes equation
up to the second-order in the Chapman-Enskog analysis. In practice,
Eqn. \ref{eq:MRT-LBE}, is implemented by first computing the post-collisional
moments $m_{i}^{*}:$
\begin{equation}
m_{i}^{*}=m_{i}-s_{i}\left(m_{i}-m_{i}^{eq}\right)+\left(I-\frac{s_{i}\delta t}{2}\right)m_{i}^{F}
\end{equation}
The post-collisional moments are then transformed back to velocity-space
distributions $f_{i}^{*}$ via $f_{i}^{*}=M^{-1}\cdot m_{i}^{*},$
after which the regular streaming step is performed. The source moments,
$m_{i}^{F}=M_{i}\cdot F_{i},$ can be pre-computed; in this work,
however, we first compute $F_{i}$ per Eqn. \ref{eq:Guo_Forcing},
and then perform a matrix-vector multiplication $\boldsymbol{M}\cdot F_{i}$
to obtain $m_{i}^{F}$. A simpler way to include force is to approximate
$F_{i}=3w_{i}\rho_{0}\frac{\boldsymbol{c}_{i}\cdot\boldsymbol{a}}{c^{2}}$
and add this term to RHS of Eqn. \ref{eq:LBE-MRT} \cite{pan2006evaluation}.
\begin{table}
\begin{centering}
\begin{tabular}{cccccccc}
\toprule 
\multirow{4}{*}{Even-order non-conserved (symmetric) moments } & \multirow{2}{*}{2nd-order } & $e$ & $p_{xx}$ & $p_{ww}$ & $p_{xy}$ & $p_{yz}$ & $p_{xz}$\tabularnewline
\cmidrule{3-8} 
 &  & $m_{1}$ & $m_{9}$ & $m_{11}$ & $m_{13}$ & $m_{14}$ & $m_{15}$\tabularnewline
\cmidrule{2-8} 
 & \multirow{2}{*}{4th-order } & $\epsilon$ & $\pi_{xx}$ & $\pi_{ww}$ &  &  & \tabularnewline
\cmidrule{3-8} 
 &  & $m_{2}$ & $m_{10}$ & $m_{12}$ &  &  & \tabularnewline
\midrule
\midrule 
\multirow{2}{*}{Odd-order non-conserved (anti-symmetric) moments} & \multirow{2}{*}{3rd-order } & $q_{x}$ & $q_{y}$ & $q_{z}$ & $m_{x}$ & $m_{x}$ & $m_{x}$\tabularnewline
\cmidrule{3-8} 
 &  & $m_{4}$ & $m_{6}$ & $m_{8}$ & $m_{16}$ & $m_{17}$ & $m_{18}$\tabularnewline
\bottomrule
\end{tabular}
\par\end{centering}
\caption{Non-conserved moments of the D3Q19 lattice \label{tab:D3Q19-Non-conserved-Moments}}
\end{table}

\subsubsection{TRT Collision Model\label{subsec:TRT-Collision-Model}}

The two-relaxation-time (TRT) model is a collision model that combines
the improved accuracy and stability of the MRT model while preserving
the simplicity of the SRT model. It was originally implemented within
the MRT framework (by assuming only two independent relaxation parameters
instead of six) \cite{pan2006evaluation}, and later was developed
independently. Unlike the MRT model, where the populations are projected
on to a space spanned by the polynomial basis vectors, the TRT collision
is based on a symmetry argument where the populations are projected
on a pair of \emph{link} basis vectors. For example, for a particular
discrete velocity $\boldsymbol{c}_{i}$ there is a anti-symmetric
velocity $\boldsymbol{c}_{\bar{i}}=-\boldsymbol{c}_{i}$, which is
true for all velocities in a particular velocity set; $\{c_{i},c_{\bar{i}}\}$
constitute a pair of link basis vector. A projection on to link basis
vectors leads to a decomposition of distributions into symmetric and
the anti-symmetric parts, denoted by a $f_{i}^{+}$and $f_{i}^{-},$
respectively, and which can be defined as \cite{kruger2017lattice}:
\begin{align}
f_{i}^{+} & \equiv\frac{f_{i}+f_{\bar{i}}}{2},\qquad f_{i}^{-}\equiv\frac{f_{i}-f_{\bar{i}}}{2}\label{eq:pdf_symm_antisymm}
\end{align}
\begin{equation}
f_{0}^{+}=f_{0},\quad f_{0}^{-}=0
\end{equation}
The symmetric and anti-symmetric part of the equilibrium distributions
$f_{i}^{eq+}$ and $f_{i}^{eq-}$, respectively, are defined similarly.
Given this decomposition, the TRT collision model can be written as:

\begin{equation}
C_{i}^{TRT}=-s^{+}\left(f_{i}^{+}-f_{i}^{eq+}\right)\delta t-s^{-}\left(f_{i}^{-}-f_{i}^{eq-}\right)\delta t
\end{equation}
and the TRT-LBE is:
\begin{equation}
f_{i}(\boldsymbol{x}+\boldsymbol{c}_{i}\Delta t,t+\Delta t)=f_{i}(\boldsymbol{x},t)-s^{+}\left(f_{i}^{+}-f_{i}^{eq+}\right)\delta t-s^{-}\left(f_{i}^{-}-f_{i}^{eq-}\right)\delta t.\label{eq:TRT-LBE}
\end{equation}
The symmetric and anti-symmetric distributions each relax with locally
prescribed relaxation rates $s^{+}$ and $s^{-}$, respectively. $s^{+}$
is linked to the fluid kinematic viscosity via $\nu=c_{s}^{2}\left(\frac{1}{s^{+}\delta t}-0.5\right)\delta t$,
and $s^{-}$ is a free parameter $s^{-}\in(0,2)$. It should be noted
that the functional form of the equilibrium distributions ($f_{i}^{eq+},\,f_{i}^{eq-}),$
the lattice weights $w_{i}$, and the lattice constant $c_{s},$ are
different from those used in the SRT and MRT collisions. For instance,
in MRT-D3Q19, the lattice constant $c_{s}=1/\sqrt{3},$ while in the
TRT model, $c_{s}$ is a tunable positive parameter. Refer to the
original references for further details \cite{d2009viscosity,ginzburg2008two,talon2012assessment}.

The TRT formulation can be also derived from the MRT collision model.
That is, the TRT collision model can be regarded as a particular form
of the MRT model when relaxation parameters associated with the even-order
non-conserved MRT moments are set to $s^{+}$ and odd-order non-conserved
moments are set to $s^{-}$ \cite{ginzburg2008two,ginzburg2003multireflection,luo2011numerics,khirevich2015coarse}
(see Tables \ref{tab:D3Q19-Non-conserved-Moments} and \ref{tab:Different-sets-of}).
Finally, when all $\{s_{i}\}=s^{+}=s^{-}=\frac{1}{\tau}$, the SRT
collision model is recovered. At second order, the incompressible
macroscopic mass and momentum conservation equations are identical
for the SRT, TRT, and MRT models, but these models differ for higher-order
approximations.

\paragraph{Implementations of TRT Model}

Practically, there are different ways in which a 'two-relaxation-time'
model can be implemented in a software code. The \emph{standard} TRT
collision model is the formulation based on the 'link' basis vectors
and which is followed by the original developers of the model, and
which was briefly discussed above. A second method of implementing
a TRT model is to employ Eqns. \ref{eq:pdf_symm_antisymm}- \ref{eq:TRT-LBE},
but rather than the equilibrium and weights developed for the standard
TRT, the usual second-order athermal equilibrium distribution of the
SRT collision, i.e., Eqn. \ref{eq:BGK-EDF} is used along with usual
weights and lattice constants of the standard D3Q19 lattice \cite{mattila2016prospect}.
The third method of implementation is based on the \emph{full} MRT
formulation, where the even-order MRT moments relax with $s^{+}$
and odd-order MRT moments relax with $s^{-}$, as identified in Table
\ref{tab:Different-sets-of}. Such a TRT implementation is equivalent
to the standard implementation, but is computationally less efficient.
A main advantage, though, is that an existing MRT code base can be
used to obtain TRT-like solutions instead of developing a new standard
TRT code. This approach can be seen as MRT in form (in terms of code)
but TRT in effect. This study adopts the third approach.

\subsection{Choice of relaxation parameters\label{subsec:Choice-of-params}}

The MRT collision model provides the most degrees of freedom to tune
relaxation rates (parameters) depending upon the flow type and configuration.
For the D3Q19 MRT model, there are six free relaxation rates, namely
$s_{e},s_{\epsilon},s_{q},s_{\pi},s_{m}$ and $s_{\nu}.$ 'Optimal'
values of these depends on particular flow type, boundary conditions,
etc., and has to be determined via parametric studies or linear stability
analysis - a task impractical for most flows of practical interests
and complexity \cite{luo2011numerics}. In the following, we list
five sets of MRT relaxation rates that were examined in this work.
To further aid clarity, we have also presented them in a tabular form
in Table \ref{tab:Different-sets-of}.
\begin{itemize}
\item \textbf{Set A}: In this case, the relaxation rates related to viscous
stresses $s_{\nu},$ are set per the kinematic viscosity, $s_{v}=\frac{1}{\tau}=\frac{2}{6\nu+1},$
and all other non-hydrodynamic or kinetic relaxation parameters are
assigned a different value. The kinetic parameters, however, are related
to the hydrodynamic ones via the relation $8\frac{(2-s_{\nu})}{(8-s_{\nu})}$.
This set of relaxation rates is the same as the 'Case A' used in \cite{pan2006evaluation}.
\item \textbf{Set B:} The relaxation rates related to bulk, kinetic-energy
squared, and viscous stresses are related and set per $s_{e}=s_{\epsilon}=s_{\pi}=s_{\nu}=\frac{2}{6\nu+1},$
and the rest of the kinetic moments relax per $s_{m}=s_{q}=8\frac{(2-s_{\nu})}{(8-s_{\nu})}$.
With this choice, the MRT model reduces to a TRT model, since one
relaxation rate is chosen for all even-ordered (symmetric) moments
and another for all odd-ordered (anti-symmetric) modes, while keeping
the specific combinations of the relaxation rates fixed when $\nu$
varies. In this case, Set B, in effect, reduces to a TRT model with
$\Lambda=3/16$ \cite{ginzburg2006variably,ginzburg2003multireflection,ginzburg2008two,luo2011numerics}.
Note that for $\Lambda=3/8,$ we have $s_{m}=s_{q}=4\frac{(2-s_{\nu})}{(4+s_{\nu})}$.
\item \textbf{Set C:} The set refers to the relaxation rates introduced
the seminal paper \cite{d2002multiple}, and is the most widely used
choice across the LBM field \cite{fattahi2016lattice,narvaez2010quantitative,saxena2017references}.
Here, $s_{\nu}$ and $s_{e}$ are independently varied, and all other
relaxation rates are set to their 'optimized' values, or set to $1$.
It should be noted that this set is 'optimal' from a stability perspective
(since $s_{e}$ controls the bulk viscosity), and was obtained by
a linear stability analysis of a 2-D non-linear shear-flow decay.
Hence, this set may not optimal for 3-D Stokes flow through porous
media.
\item \textbf{Set D}: In this case, similar to Set B, the relaxation rates
of even-ordered moments are related and set per $s_{e}=s_{\epsilon}=s_{\pi}=s_{\nu}=\frac{2}{6\nu+1}.$
However, we set the rates for the two third-order moments $s_{q}$
and $s_{m}$ independently: $s_{q}=8\frac{(2-s_{\nu})}{(8-s_{\nu})}$
and $s_{m}=4\frac{(2-s_{\nu})}{(4+s_{\nu})}$. The set is obtained
from the parameterization rule (Rule (19) in \cite{khirevich2015coarse})
by setting two separate values for the third-order (anti-symmetric)
moments, i.e., \textcolor{black}{$\Lambda_{q}=\big(\frac{1}{s_{q}}-\frac{1}{2}\big)=3/16$
and $\Lambda_{m}=\big(\frac{1}{s_{m}}-\frac{1}{2}\big)=3/8.$ }This
set effectively reduces the full MRT model to a three-relaxation time
(TrRT) with three independent parameters $s_{\nu}$, $s_{q}$ and
$s_{m}$, but, crucially, where $s_{m}$ and $s_{q}$ vary with $s_{\nu}$. 
\item \textbf{Set E:} This set is a generalization of Set B, by which one
can obtain TRT-like solutions using a MRT framework. Therefore, only
two independent parameters are to be chosen. As mentioned before,
all even-order moments relax with $s^{+}=s_{\nu},$ which controls
kinematic viscosity. Odd-ordered moments relax with $s^{-}$ but such
that a related parameter remains constant. This parameter, called
the 'magic' parameter $\Lambda,$ is defined as:

\begin{equation}
\Lambda=\text{\ensuremath{\Lambda}}^{+}\Lambda^{-}=\bigg(\frac{1}{s^{+}}-\frac{1}{2}\bigg)\bigg(\frac{1}{s^{-}}-\frac{1}{2}\bigg)\label{eq:Magic param}
\end{equation}
Thus, instead of $s^{+}$and $s^{-}$ being the two independent parameters,
in practice, $s^{+}$ and $\Lambda$ are the independent parameters.
That is, for a chosen $\Lambda,$ if $s^{+}$ is changed to vary the
viscosity, then $s^{-}$ also needs to be changed such that $\Lambda$
is kept constant. A question therefore arises as to how to choose
an optimal value of $\Lambda$. In terms of accuracy and stability,
certain \emph{special} values of $\Lambda$ show distinctive proprieties
depending upon the flow type and configuration. More specifically,
an analysis of LBM behavior for Poiseuille flow previously showed
that $\Lambda$ controls the location of the no-slip (zero-velocity)
boundary between voxels tagged as fluid and solid. Through such an
analysis, several specific optimal values of $\Lambda$ have been
obtained for simple geometries that allow exact analytical solutions.
For example, using the bounceback scheme and for Poiseuille flow in
a horizontal straight channel, a value of $\Lambda=3/16$ results
in the solid boundary being located exactly in the middle of the solid
and fluid voxel, i.e., $\delta x/2$ beyond the last fluid voxel.
It also yields the exact permeability \cite{talon2012assessment}.
Similarly, with the bounceback scheme and Poiseuille flow in a diagonal
channel, a value of $\Lambda=3/8$ was obtained. However, for Stokes
flow through arbitrary porous geometry and using the bounceback rule,
no unique values for $\Lambda$ exists \cite{d2009viscosity,khirevich2015coarse}.
Therefore, in order to understand the variance of solutions with $\Lambda,$
we use the two 'basic' values of $\Lambda=\{\frac{3}{16},\frac{3}{8}\},$
in addition to $\Lambda=1/64$ which represents a 'small' value. Also
note that $\Lambda=1/4$ is equivalent to the SRT model with $\tau=1,$
and for $\Lambda=3/16$ the relaxation rates corresponds to those
in Set B.
\end{itemize}
Finally, it should be noted that if all six relaxation rates that
are afforded in the D3Q19 MRT model need to be independently set,
then the relaxation rates have to follow the parameterization rule
described as Rule 19 in \cite{khirevich2015coarse} in order to obtain
viscosity-independent results for Stokes flow.
\begin{center}
\begin{table}
\caption{Different sets of MRT relaxation rates for the D3Q19 lattice.\label{tab:Different-sets-of}}
\centering{}%
\begin{tabular}{>{\centering}p{0.5in}c>{\centering}p{0.9in}>{\centering}p{0.9in}>{\centering}p{0.9in}>{\centering}p{0.9in}>{\centering}p{0.9in}}
\toprule 
Moment & Number & Set A & Set B & Set C & Set D  & Set E\tabularnewline
\midrule
\midrule 
$s_{\rho}$ & $s_{0}$ & $0$ & $0$ & $0$ & $0$ & $0$\tabularnewline
\midrule
\midrule 
$s_{e}$ & $s_{1}$ & $8\frac{(2-s_{\nu})}{(8-s_{\nu)}}$ & $s_{\nu}=\frac{1}{\tau}$ & $1.19$  & $s_{\nu}=\frac{1}{\tau}$ & $s_{\nu}=\frac{1}{\tau}$\tabularnewline
\midrule
\midrule 
$s_{\epsilon}$ & $s_{2}$ & $8\frac{(2-s_{\nu})}{(8-s_{\nu)}}$ & $s_{\nu}=\frac{1}{\tau}$ & $1.4$ & $s_{\nu}=\frac{1}{\tau}$ & $s_{\nu}=\frac{1}{\tau}$\tabularnewline
\midrule
\midrule 
$s_{j_{x}}$ & $s_{3}$ & $0$ & $0$ & 0 & $0$ & $0$\tabularnewline
\midrule
\midrule 
$s_{q_{x}}$ & $s_{4}$ & $8\frac{(2-s_{\nu})}{(8-s_{\nu)}}$ & $8\frac{(2-s_{\nu})}{(8-s_{\nu)}}$ & $1.2$ & $8\frac{(2-s_{\nu})}{(8-s_{\nu)}}$ & $2\frac{(2-s_{\nu})}{(4\Lambda s_{\nu}-s_{\nu}+2)}$\tabularnewline
\midrule
\midrule 
$s_{j_{y}}$ & $s_{5}$ & $0$ & $0$ & $0$ & $0$ & $0$\tabularnewline
\midrule
\midrule 
$s_{q_{y}}$ & $s_{6}$ & $8\frac{(2-s_{\nu})}{(8-s_{\nu)}}$ & $8\frac{(2-s_{\nu})}{(8-s_{\nu)}}$ & $1.2$ & $8\frac{(2-s_{\nu})}{(8-s_{\nu)}}$ & $2\frac{(2-s_{\nu})}{(4\Lambda s_{\nu}-s_{\nu}+2)}$\tabularnewline
\midrule
\midrule 
$s_{j_{z}}$ & $s_{7}$ & $0$ & $0$ & 0 & $0$ & $0$\tabularnewline
\midrule
\midrule 
$s_{q_{z}}$ & $s_{8}$ & $8\frac{(2-s_{\nu})}{(8-s_{\nu)}}$ & $8\frac{(2-s_{\nu})}{(8-s_{\nu)}}$ & $1.2$ & $8\frac{(2-s_{\nu})}{(8-s_{\nu)}}$ & $2\frac{(2-s_{\nu})}{(4\Lambda s_{\nu}-s_{\nu}+2)}$\tabularnewline
\midrule
\midrule 
$s_{\nu}$ & $s_{9}$ & $s_{\nu}=\frac{1}{\tau}$ & $s_{\nu}=\frac{1}{\tau}$ & $s_{\nu}=\frac{1}{\tau}$  & $s_{\nu}=\frac{1}{\tau}$ & $s_{\nu}=\frac{1}{\tau}$\tabularnewline
\midrule
\midrule 
$s_{\pi}$ & $s_{10}$ & $8\frac{(2-s_{\nu})}{(8-s_{\nu)}}$ & $s_{\nu}=\frac{1}{\tau}$ & $1.4$ & $s_{\nu}=\frac{1}{\tau}$ & $s_{\nu}=\frac{1}{\tau}$\tabularnewline
\midrule
\midrule 
$s_{\nu}$ & $s_{11}$ & $s_{\nu}=\frac{1}{\tau}$ & $s_{\nu}=\frac{1}{\tau}$ & $s_{\nu}=\frac{1}{\tau}$ & $s_{\nu}=\frac{1}{\tau}$ & $s_{\nu}=\frac{1}{\tau}$\tabularnewline
\midrule
\midrule 
$s_{\pi}$ & $s_{12}$ & $8\frac{(2-s_{\nu})}{(8-s_{\nu)}}$ & $s_{\nu}=\frac{1}{\tau}$ & $1.4$ & $s_{\nu}=\frac{1}{\tau}$ & $s_{\nu}=\frac{1}{\tau}$\tabularnewline
\midrule
\midrule 
$s_{\nu}$ & $s_{13}$ & $s_{\nu}=\frac{1}{\tau}$ & $s_{\nu}=\frac{1}{\tau}$ & $s_{\nu}=\frac{1}{\tau}$ & $s_{\nu}=\frac{1}{\tau}$ & $s_{\nu}=\frac{1}{\tau}$\tabularnewline
\midrule
\midrule 
$s_{\nu}$ & $s_{14}$ & $s_{\nu}=\frac{1}{\tau}$ & $s_{\nu}=\frac{1}{\tau}$ & $s_{\nu}=\frac{1}{\tau}$ & $s_{\nu}=\frac{1}{\tau}$ & $s_{\nu}=\frac{1}{\tau}$\tabularnewline
\midrule
\midrule 
$s_{\nu}$ & $s_{15}$ & $s_{\nu}=\frac{1}{\tau}$ & $s_{\nu}=\frac{1}{\tau}$ & $s_{\nu}=\frac{1}{\tau}$ & $s_{\nu}=\frac{1}{\tau}$ & $s_{\nu}=\frac{1}{\tau}$\tabularnewline
\midrule
\midrule 
$s_{m}$ & $s_{16}$ & $8\frac{(2-s_{\nu})}{(8-s_{\nu)}}$ & $8\frac{(2-s_{\nu})}{(8-s_{\nu)}}$ & $1.98$ & $4\frac{(2-s_{\nu})}{(4+s_{\nu)}}$ & $2\frac{(2-s_{\nu})}{(4\Lambda s_{\nu}-s_{\nu}+2)}$\tabularnewline
\midrule
\midrule 
$s_{m}$ & $s_{17}$ & $8\frac{(2-s_{\nu})}{(8-s_{\nu)}}$ & $8\frac{(2-s_{\nu})}{(8-s_{\nu)}}$ & $1.98$ & $4\frac{(2-s_{\nu})}{(4+s_{\nu)}}$ & $2\frac{(2-s_{\nu})}{(4\Lambda s_{\nu}-s_{\nu}+2)}$\tabularnewline
\midrule
\midrule 
$s_{m}$ & $s_{18}$ & $8\frac{(2-s_{\nu})}{(8-s_{\nu)}}$ & $8\frac{(2-s_{\nu})}{(8-s_{\nu)}}$ & $1.98$ & $4\frac{(2-s_{\nu})}{(4+s_{\nu)}}$ & $2\frac{(2-s_{\nu})}{(4\Lambda s_{\nu}-s_{\nu}+2)}$\tabularnewline
\bottomrule
\end{tabular}
\end{table}
\par\end{center}

\subsection{Inlet/outlet Boundary Conditions\label{subsec:Boundary-Conditions}}

Two types of boundary conditions are required to simulate flow through
a porous medium: no-slip boundary condition on the voxels labeled
as solid/obstacle/matrix in the 3-D image, and boundary conditions
at the inlet/outlet and lateral boundaries. For enforcing the no-slip
boundary condition on solid voxels, we apply the so-called \emph{full-way}
bounceback scheme, an implementational variant of the bounceback rule.
In the full-way bounceback implementation, the distributions are assumed
to travel the complete distance from the bounceback voxel to the solid
voxel, where the velocity is reversed in the collision of the next
timestep/iteration. The full-way bounceback is simple to implement,
and robust for tortuous pore geometries, as compared to half-way bounceback.

The choice of boundary conditions on the inlet and outlet planes and
on lateral planes (planes normal to the flow direction) depends on
the method used to drive the flow. There are two ways to drive the
flow. The most widely used method is to apply a uniform body force,
$\boldsymbol{F}=\rho\boldsymbol{a},$ at every node of the domain,
which corresponds to the presence of gravity or pressure gradient
via $\nabla p=\frac{P_{in}-P_{out}}{L}=\rho\boldsymbol{a}$ \cite{eshghinejadfard2016calculation,fattahi2016lattice,mattila2016prospect,saxena2017references}.
The force is then incorporated in the LBE equation using Eqn. \ref{eq:MRT-LBE}.
The use of body forcing, however, requires that periodic boundary
conditions be imposed on the inlet and outlet planes. Physically,
for a computational domain boundary to be periodic, the mass flux
exiting a plane perpendicular to the boundary (e.g. outlet) has to
re-enter through the opposite (inlet) plane, i.e., the pore/solid
topology at the inlet and outlet boundaries have to match exactly.
While this condition might be satisfied for some porous media models,
such as body-centered cubic (BCC) and face-centered cubic (FCC) sphere
packing, it is strictly not valid for actual rocks and stochastic
porous media samples. Application of periodic boundary conditions
in these case may result in creation and destruction of flow pathways
resulting in increased/decreased permeability predictions. To obtain
a truly periodic domain, one has to mirror the porous medium in each
of the coordinates axes, which leads to an 8X increase in the computational
domain size \cite{fredrich2006predicting}.

To address this issue, in this work (in Sec. \ref{subsec:Sphere-Pack}),
we use the technique of padding the inlet and outlet planes (i.e.,
planes perpendicular to the direction of applied forcing) with 5 lattice-thick
layer of fluid-only cells. This technique is variously referred to
as padding/jacketing/buffering \cite{Ferrol1995,fredrich2006predicting,narvaez2010quantitative,manwart2002lattice}..
Additionally, on the planes perpendicular to the direction of the
flow, we apply the symmetry/mirror boundary condition. In a symmetric
boundary condition, the mass flux towards the boundary from one side
must be accompanied by a mirrored flow from the other side resulting
in zero net mass flux perpendicular to the symmetry boundary. Macroscopically,
the symmetry boundary implies at the symmetry plane we have zero normal
velocity and zero normal gradients of all variables. LB implementation
is straightforward: after the collision step a mirror image of all
particle distributions is constructed for each voxel that borders
the boundary. The second method to drive the flow is by applying a
pressure gradient across the sample by prescribing pressure at the
inlet and at the outlet. This technique is used in N-S-based solvers.
In LBM, however, due to the ideal gas $(p=\rho c_{s}^{2})$ and incompressible
flow assumptions, the pressure gradient is implemented by applying
a density difference across the sample. Commonly used techniques to
apply a prescribed value of density at inlet or outlet boundaries
include non-equilibrium bounceback (Zou-He), non-equilibrium extrapolation,
(GZS), and the anti-bounceback scheme. Refer to \cite{guo2013lattice,kruger2017lattice}
for a detailed discussion on the above- mentioned schemes.

\section{Numerical Tests\label{sec:Numerical-Tests}}

\begin{figure}
\centering{}\subfloat[Channel with low-resolution circular cross-section \label{fig:lowRes}]{\begin{centering}
\includegraphics{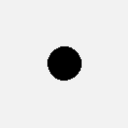}
\par\end{centering}

}\qquad{}\subfloat[Channel with high- resolution circular cross-section \label{fig:HighRes}]{\begin{centering}
\includegraphics[scale=0.55]{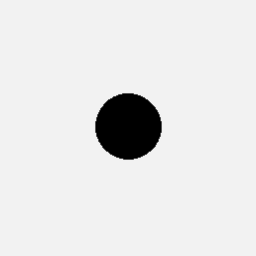}
\par\end{centering}
}\qquad{}\subfloat[Channel with a triangular cross-section \label{fig:TriCS}]{\begin{centering}
\includegraphics[scale=0.55]{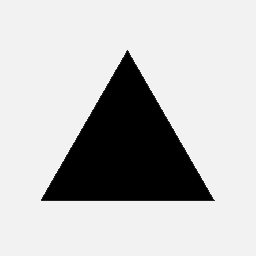}
\par\end{centering}

}\caption{Model pore geometries. Here BLACK indicate pore space available for
the fluid phase and WHITE indicates solid or matrix phase \label{fig:Model-pore-geometries}}
\end{figure}

In this section, we discuss the application of the collision models
and the effects of choosing different sets of relaxation parameters,
as described in the previous section. The primary property of interest
is the intrinsic (absolute) permeability. Three geometric models of
porous media were considered for which theoretical or numerical results
are available: (i) a 3-D square block permeated with a channel of
uniform circular cross-section (Figs. \ref{fig:lowRes} and \ref{fig:HighRes});
(ii) a 3-D square channel permeated with a channel of uniform triangular
cross-section (Fig. \ref{fig:TriCS}); (iii) a 3-D random pack of
identical spheres. For the first two cases, the images representing
the cross-sections were created using \emph{Paint.net}, a graphics
editor program, and then combined into a single 3-D image stack using
\emph{ImageJ}. The 3-D image stack for the pack of spheres was obtained
from supplementary material provided by \cite{saxena2017references}
and which is also available at \cite{Finney}. The images were analyzed
and converted into a .dat file using a \emph{MATLAB} script which
serves as an input to the LBM simulator. In the image-analysis phase,
each voxel in the stack is categorized as a \texttt{fluid} voxel if
all its 18 neighbors are \texttt{fluid} and then assigned a flag value
of \texttt{0}, a \texttt{bounceback} voxel if it has both \texttt{solid}
and \texttt{fluid} neighbors and is assigned a flag value 1, and a
\texttt{solid} voxel if all its neighbors are \texttt{solid} voxels
and then assigned a flag value of \texttt{2}. The simulator is developed
using \emph{Palabos}, a open-source, parallel, LBM software \cite{Palabos}.
From an LBM implementation perspective, normal streaming and collision
is performed at a \texttt{fluid} voxel, no LB operations are performed
at a \texttt{solid} voxel, and the local collision rule is modified
so as to reverse the directions of the incoming populations at a \texttt{bounceback}
voxel.

The main output of the LBM simulator is the local velocity field,
$\boldsymbol{u}(\boldsymbol{x}),$ which is the steady-state solution
of the Navier-Stokes equations as recovered by the collision models.
Based on Darcy's law, the diagonal component of the permeability tensor
$\boldsymbol{K}$ in the $j$th direction is the computed as \cite{talon2012assessment}:
\begin{equation}
k_{j}=K_{jj}=\frac{\nu\langle u_{j}\rangle}{\langle\nabla p-\rho_{0}F_{j}\rangle}\label{eq:perm_calc}
\end{equation}
where $\langle u_{j}\rangle$ is the volume average of the $j$th
component of velocity field and $\langle\nabla p-\rho_{0}F_{j}\rangle$
is the component of the driving force in the $j$ direction. When
an external body force is not present, $\nabla p=\frac{\langle p_{in}\rangle-\langle p_{out}\rangle}{Nx-1}$,
where $\langle p_{in}\rangle$and $\langle p_{out}\rangle$ represents
the average pressure at the inlet and outlet boundaries, respectively,
and $Nx$ is the number of the images in the direction of forcing.
When forcing is used, $\nabla p$ is neglected in the calculation
for $k_{j}$. For all of the results presented below, a steady-state
solution is assumed when the normalized difference between permeability
values between time $n$ and $n-1$ is less than $10^{-6}$:
\begin{equation}
\left|\frac{k_{j}^{n}-k_{j}^{n-1}}{k_{j}}\right|<10^{-6}.\label{eq:convergence}
\end{equation}

\subsection{Channel with a Circular Cross-Section\label{subsec:Circular-Cylinder}}

Perhaps the simplest porous media model that can be used for validation
purposes is a 3-D channel permeated with a cylindrical bore. Although
this geometric model does not have pores and pore throats like a real
rock, it nevertheless serves as a useful means to test the effect
of relaxation parameters in isolation without the complexity of real
porous media. It is also broadly useful to validate the implementation
aspects of the simulator, such as boundary conditions, pre-processing
methods, etc. Figs. \ref{fig:Model-pore-geometries} (a) and (b) show
the geometry of the model. Combining the analytical solution for laminar
Poiseuille flow in a pipe and Darcy's law, we can obtain the theoretical
(reference) permeability for a channel with a cylindrical bore \cite{mavko2009rock}:
\begin{equation}
k_{t}^{c}=\frac{\pi R^{4}}{8A}\label{eq:k_th_circleCS}
\end{equation}
where $A$ is the cross-sectional area of channel. In the simulations
presented in this sub-section, we prescribe Dirichlet boundary conditions
for density (pressure) at the inlet and outlet using the Zou-He boundary
scheme. The pressure difference is kept low enough that Stokes flow
$(Re\ll1)$ is maintained and LBM-compressibility errors are negligible. 

\begin{figure}
\begin{centering}
\includegraphics[scale=0.65]{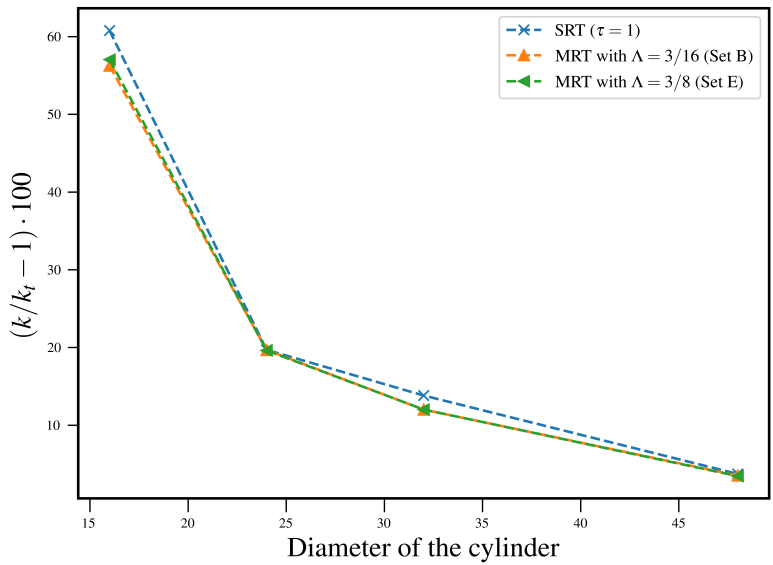}
\par\end{centering}
\caption{Errors in permeability of a circular channel as a function of the
diameter of the channel for different relaxation parameters. \label{fig:permvsporosity}}

\end{figure}

For preliminary validation of the simulator, we first compute the
permeability of a channel as a function of its diameter (porosity).
A square channel of size 64x64x64 voxels is permeated with uniform
cylinders of varying diameters, $D=\{16,24,36,48\}$ in lattice units.
Since this study focuses on pore-spaces obtained from microtomographic
images, this test models a hypothetical pore which is being resolved
with higher resolution with increasing diameter. Various permeabilities
are computed with the SRT model with $\tau=1$, MRT models with the
Set B parameters and $s_{\nu}=1,$ and the Set E parameters with $s_{\nu}=1,\,\Lambda=3/8$.
Fig. \ref{fig:permvsporosity} shows the errors in the computed permeability
as a function of diameter (porosity), where the errors are calculated
based on reference values obtained from Eqn. \ref{eq:k_th_circleCS}.
We can observe that when the pore-space (of the bore) is resolved
very coarsely, the errors in computed permeability are around $60\%$,
irrespective of the collision models and their relaxation parameters.
The errors, however, decrease from 60\% to around 3\% with increasing
diameter. The decreases in errors are clearly due to resolution effects,
since a larger diameter resolves a circular solid boundary more finely
when represented in terms of voxels. For low-porosity samples in particular,
this means that unless the pores are sufficiently resolved, any set
of MRT parameters in particular would yield relatively large errors.
This indicates that the resolution of the microtomographic images
of the pore-space is the dominant factor affecting the accuracy of
permeability predictions. This is expected since the stepped nature
of cubic voxels, in general, leads to overestimation of surface area.
For example, for a sphere approximated using cubic voxels, when the
ratio of voxel length over the radius of the sphere tends to zero
(increasingly smaller voxel length), the ratio of surface area of
a sphere modeled with voxels to that of a perfect sphere will tend
to $3/\sqrt{3}\approx1.73.$ This means that for a sphere approximated
with cubic voxels, the surface area is overestimated by around 70\%.

\begin{figure}[p]
\begin{centering}
\includegraphics{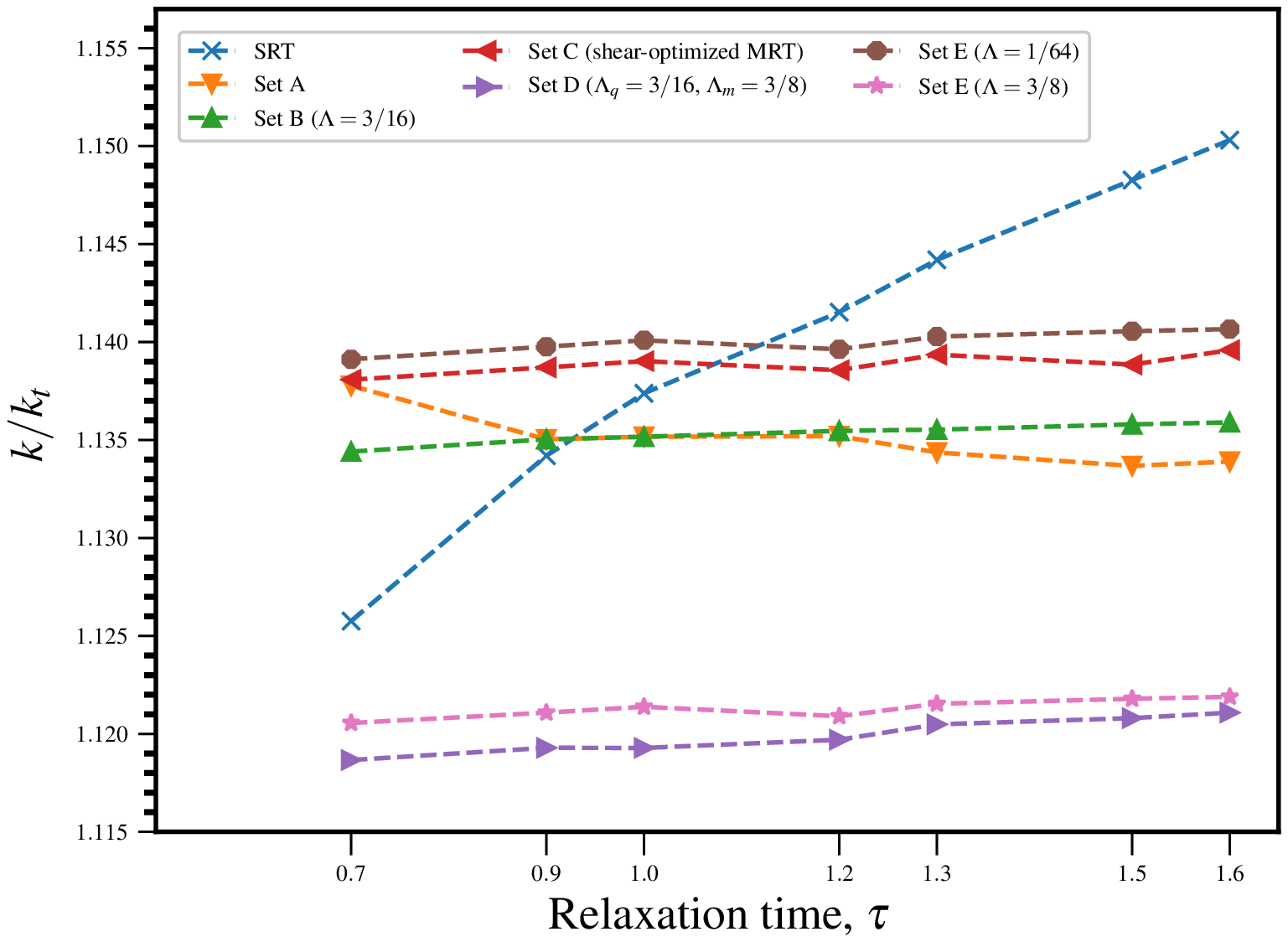}
\par\end{centering}
\caption{Normalized permeability $k/k_{ref}$ as a function of relaxation time
for channel of size $128\times128\times128$ permeated by a circular
pipe of diameter $D=32$ (low-resolution geometry).\label{fig:CircleLowRes}}

\end{figure}

\begin{figure}[p]
\begin{centering}
\includegraphics{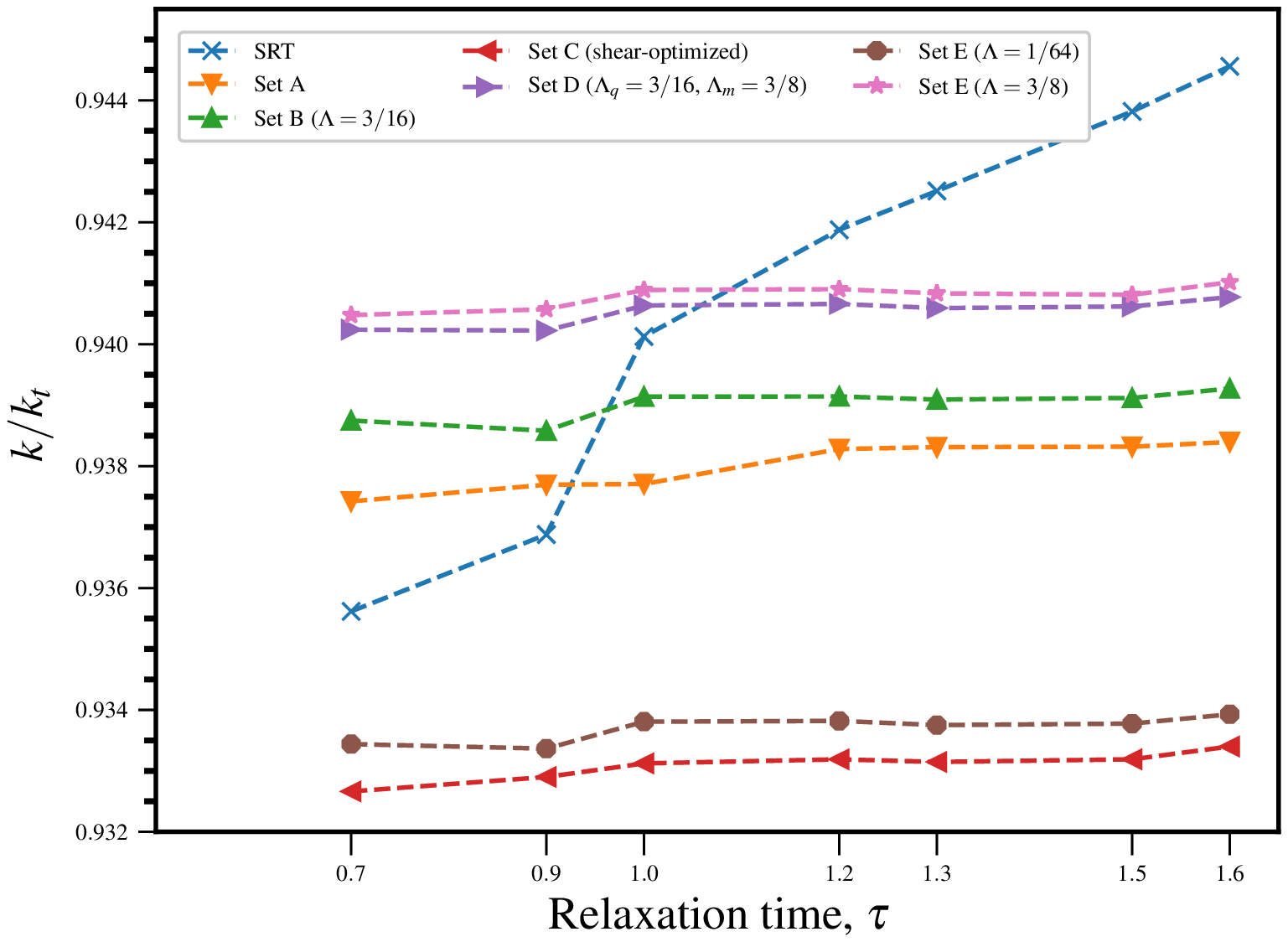}
\par\end{centering}
\caption{Normalized permeability$k/k_{ref}$ as a function of relaxation time
for channel of size $256\times256\times256$ permeated by a circular
pipe of diameter $D=64$ (medium-resolution geometry). \label{fig:CircleMedRes}}
\end{figure}

Next, we investigate the dependence of permeability predictions on
the choices of relaxation parameters within the MRT collision model.
For this purpose, we use three geometries with circular cross-sections:
(i) a low-resolution channel where the diameter of the cylinder is
32 lattice units, (ii) a medium-resolution channel where the diameter
is 64 lattice units, and (iii) a high-resolution channel where the
diameter is 128 lattice units. The channel square cross section for
the low, medium and high resolutions is $128\times128,$ $256\times256,$
and $512\times512,$ respectively, such that the porosity of all of
the geometries is maintained at $\phi=0.196.$ For each geometry,
the viscosity is varied via the relation time, $\tau=\frac{1}{s_{\nu}}$
and the rest of the MRT relaxation parameters are varied as per Tab.
\ref{tab:Different-sets-of}. Figs. \ref{fig:CircleLowRes}, \ref{fig:CircleMedRes},
and \ref{fig:CircleMedRes} show the results of the simulations on
the low, medium, and high-resolution geometries, respectively, where
normalized permeability predictions are plotted against the relaxation
time (viscosity).
 
\begin{figure}[p]
\begin{centering}
\includegraphics{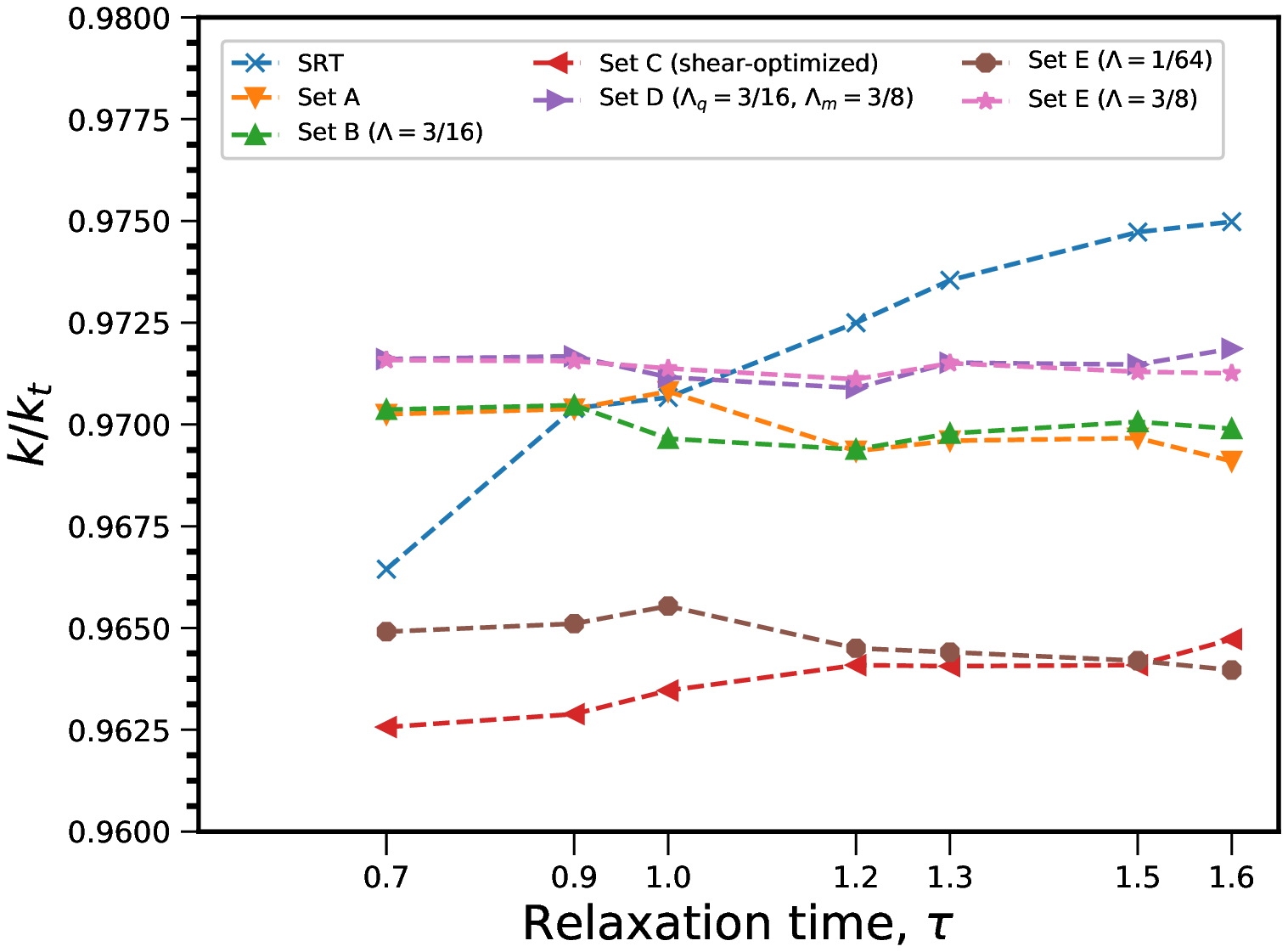}
\par\end{centering}
\caption{Errors in permeability as a function of relaxation time for channel
of size $512\times512\times512$ permeated by a circular pipe of diameter
$D=128$ (high-resolution geometry). \label{fig:CircleHighRes}}
\end{figure}

We first discuss the issue of viscosity independence. In the case
of the channel of low-resolution, the errors are relatively high $(>12\%)$
for all sets of MRT parameters, as seen in Fig. \ref{fig:CircleLowRes}.
However, \textcolor{black}{even such a coarse resolution, we can observe
that variance of the normalized permeability over the range of $\tau$
$(\nu)$ as obtained with Sets B, D, and E are much smaller $(<0.2\%)$
compared to Set A and C, where the variance is about $0.5\%$ . This
indicates that these parameters sets result in viscosity-independent
permeability values. On the other hand, there is a much wider variance
in errors for the SRT and Set A parameters. Similar trends can be
seen in case of medium-resolution images. Notably, we can observe
that the popular set of MRT parameters, i.e. Set C, has a slightly
higher variance compared to Sets B, D, and E. For both the low and
medium resolution cases,} results obtained for SRT, and Sets A and
B is consistent with the theoretical analysis which show that these
cases do not follow rules of parameterization required to produce
$\nu$-independent solutions of Stokes flow, see Rule (19) in \cite{khirevich2015coarse}.
For the high-resolution geometry, for all of the cases (except the
SRT), the variance in computed permeability is much smaller compared
to low and medium resolutions. For Sets B, D, and E, we observe error
variance of less than 0.1\% as compared to variance of about 0.3\%
for Sets A and C. We can also observe that the popular choice of parameters
(Set C) shows slightly larger variation in permeability compared to
Sets B, D and E, for both high and low resolution cases. Finally,
across all resolutions, parameter sets B, D, and E consistently lead
to $\nu-$independent results, and in general, dependence of permeability
on $\nu$ reduces with increasing resolution of the pore-space.

Next, we consider the issue of accuracy. In low-resolution geometries
we can observe that the errors are considerably higher irrespective
of collision model and their parameters. Thus, as mentioned before,
for under-resolved cases, all collision models (including MRT) can
experience large errors irrespective of parameters. In parameter Sets
B, D, and E, these errors remain fairly constant with variation of
$\tau\,(\nu)$. For Set E, we observe that as $\Lambda$ increases
from $1/64$ to $3/8,$ the accuracy increases for all three resolutions;
thus $\Lambda=3/8$ leads to the smallest errors for each of three
resolutions. Notably, in the case of low-resolution images, small
values of $\Lambda$ result in much larger errors even compared to
SRT. These trends indicate that, for a general porous medium, permeability
obtained with the bounceback rule still depends on $\Lambda$ mainly
due to unresolved pores. Moreover, \textcolor{black}{spatial truncation
error remain constant irrespective of $\tau$ and hence the permeability
remains independent of $\tau\,(\nu)$.} 

Finally, we have observed that increasing $\tau$ leads to faster
convergence for all cases, including for the SRT model. For instance,
for the high resolution case with Set D parameters, the number of
iterations required to converge decreased consistently from 19200
for $\tau=0.7$ to 7900 for $\tau=1.6.$ Similar trends were observed
for all MRT sets of parameters (A-E) and all resolutions.

\subsection{Channel with Triangular Cross-Section\label{subsec:Channel-with-Triangular}}

The next model of porous media that we investigate is a channel with
an opening of a triangular cross-section. Figure \ref{fig:TriCS}
show the details of this geometry. The reference permeability, $k_{ref}^{t},$
is computed as \cite{mavko2009rock}:
\begin{equation}
k_{ref}^{t}=\frac{\sqrt{3}l^{4}}{320A}
\end{equation}
where $l$ is the length of each side of the triangle and $A$ is
cross-sectional area of the channel. The geometry creation and analysis
procedure remains the same as described at the beginning of Sec. \ref{sec:Numerical-Tests}.
In the results that follow, we considered a channel of size $256\times256\times256$
permeated with a triangle where each side measures 200 lattice units.
\begin{figure}
\begin{centering}
\includegraphics{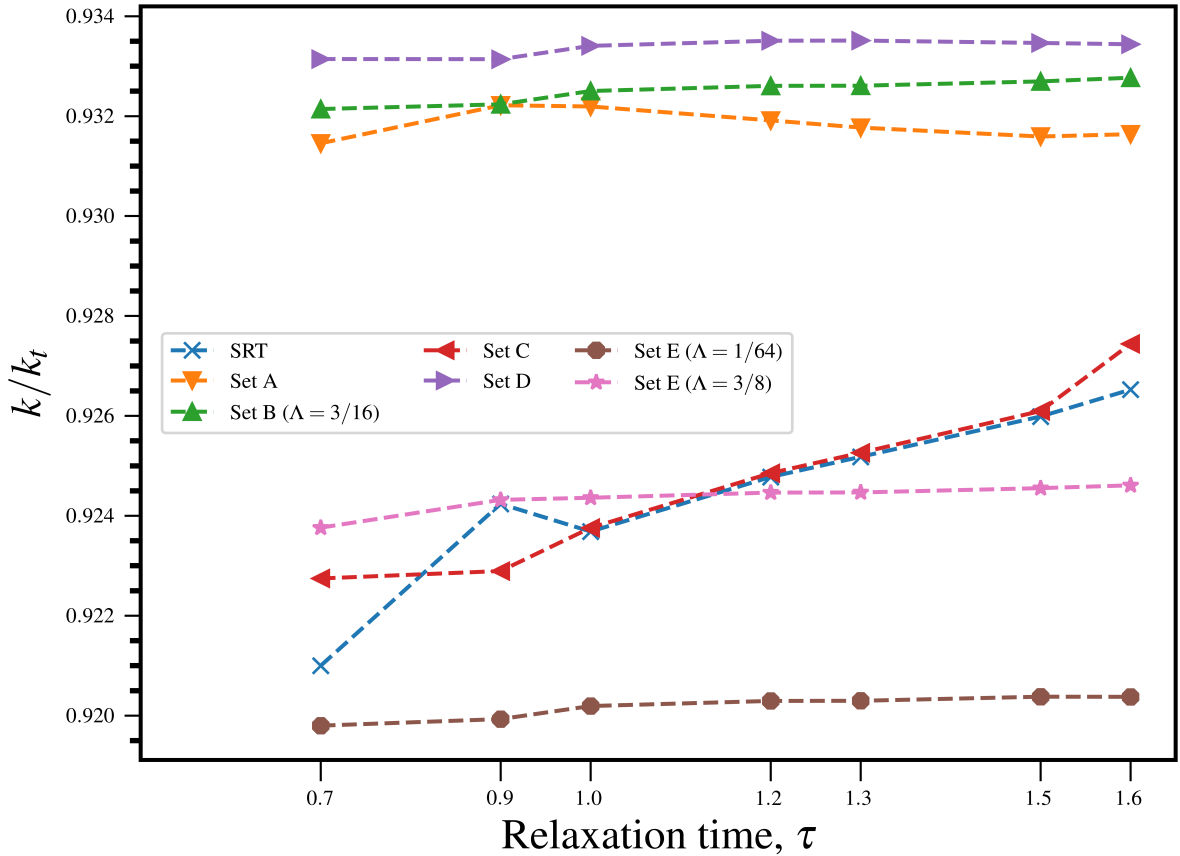}
\par\end{centering}
\caption{Normalized permeability function of relaxation time for channel of
size $256\times256\times256$ permeated by a triangular cross-section
of side length $l=200$ and $L=256$\label{fig:TriangleCS}}
\end{figure}

In this case of a relatively high-resolution geometry, we can observe
from Fig. \ref{fig:TriangleCS}, that parameter Sets D and E result
in very low variance (less than 0.1 \% in each case) in permeability
prediction over the entire range of $\tau$ that were tested. Similar
to the case of circular cross-section, we can observe that with Set
C parameters, the variance is much stronger than in Sets D and E.
Among the three sets of parameters, we find that Set D, where $\Lambda_{q}=3/16$
and $\Lambda_{m}=3/8,$ gives the most accurate overall predictions
for the channel with triangular cross-section. We also observe that
$k$ varies with $\Lambda,$ which indicates that overall accuracy
of permeability predictions still depends on $\Lambda.$ This is,
again, due to the finite-resolution of the images where solid boundary
and pore-spaces may not be sufficiently resolved.

\subsection{Sphere Pack\label{subsec:Sphere-Pack}}

\begin{figure}
\centering{}\subfloat[Segmented image of a cross-section. Here BLACK indicate pore space
and WHITE indicates solid or matrix space\label{fig:Segmented-image-of}]{\begin{centering}
\includegraphics[scale=0.5]{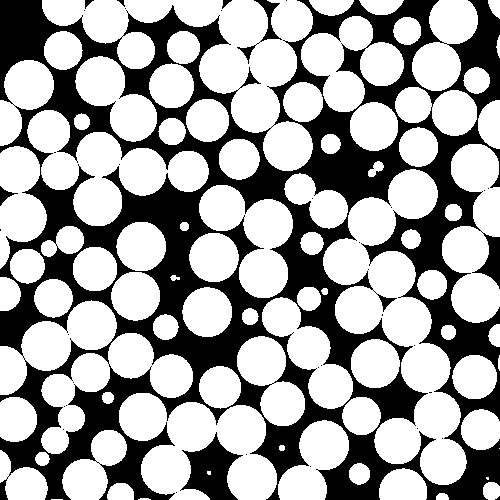}
\par\end{centering}
}\qquad{}\subfloat[3-D rendering of the pore-space in random packing of spheres, where
the gray voxel represent the pore/fluid space. Voxel resolution is
$7\,\mu m$ and porosity of the packing is 0.34.\label{fig:3-D-visualization-of} ]{\begin{centering}
\includegraphics[scale=0.55]{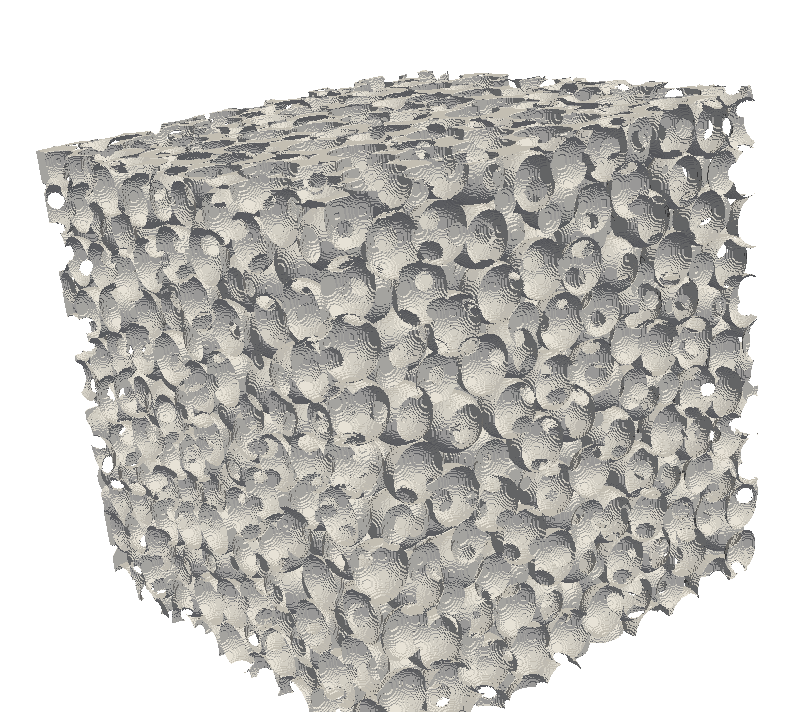}
\par\end{centering}
}\qquad{}\caption{Random packing of mono-disperse spheres \label{fig:Model-pore-geometries-1}}
\end{figure}

Random dense packing of identical (monodisperse) spheres serves as
a useful geometry for the validation of pore-scale simulations for
several reasons. First, experimental measurements can be easily obtained
with sintered glass beads, and both empirical correlations and laboratory
measurements are widely available. Second, random sphere packs are
reasonable first-order approximations to sandstone, and yet are well
defined mathematically. Third, a wide range of porosity and resolution
can be obtained both in the laboratory and in simulations by uniform
growth or shrinking of the spheres of the packing. In this study,
we use the random sphere packing used in \cite{saxena2017references};
the entire image dataset is available online at \cite{BenchmarkDatabase}.
The sphere pack dataset consists of 793 .\texttt{bmp} images, each
of size 788x791. Fig. \ref{fig:Segmented-image-of} shows a cross
sectional view (in the direction of applied forcing) of the sphere
pack and Fig. \ref{fig:3-D-visualization-of} shows a 3-D rendering
of the geometry. The diameter of the spheres is 0.714 mm, and the
voxel resolution is 7$\mu m$, giving a lattice nodes per sphere diameter,$\,d_{sp}\approx100.$
The porosity of the sample is $\phi=0.34.$ We again point out that,
in this work, we do not use curved boundary conditions to represent
the surface of the spheres; instead, we use the bounceback boundary
condition which the method of choice for pore-spaces obtained from
microtomographic images.

As before, the images were converted to a .\texttt{dat} file which
serves as an input for the LBM simulator. We apply the symmetry boundary
condition on domain boundaries that are normal to the flow directions.
Along the flow direction, a 5-layer padding is applied each at the
inlet and outlet domain, as described in Sec. \ref{subsec:Boundary-Conditions}.
The flow is imposed by applying a uniform body force in the positive
$x$-direction: $F_{x}=\rho_{o}g_{x}$ with $\rho_{o}=1$ and $g_{x}=0.0001;$
no pressure gradient is explicitly imposed. From the steady-state
solution (Stokes velocity field), the permeability is computed using
Eqn. \ref{eq:perm_calc}. For reference values, we use a value of
$k_{ref}^{sp}=280,151mD$ which is the median value of permeability
that was obtained using different LBM and non-LBM based fluid solvers
on the same sphere pack image datasets as used in this study \cite{saxena2017references}.
The reference sphere pack permeability translates to $k_{ref,lb}^{sp}=5.62$
l.u if expressed in LB units. Permeability in Darcy units can be converted
to LB units, and vice-versa, via the relation $k_{D}$ =$k_{lb}\times\Delta x^{2}\times0.9869,$
where $k_{D}$ is permeability in Darcy and $\Delta x$ is the voxel
resolution in $m$.

The simulations are performed at three values of relaxation time,
$\tau=\{0.7,1.0,1.5\}.$ Moreover, in addition to previously tested
parameter sets, we also performed simulations with Set E parameters
with two additional values of $\Lambda$, namely $\Lambda=1/64$ and
$\Lambda=1,$ which represent a lower and higher bound on $\Lambda$
values as compared to the 'basic' values of $\{\frac{1}{8},\frac{3}{16},\frac{1}{4},\frac{3}{8}\}.$

\begin{figure}
\begin{centering}
\includegraphics{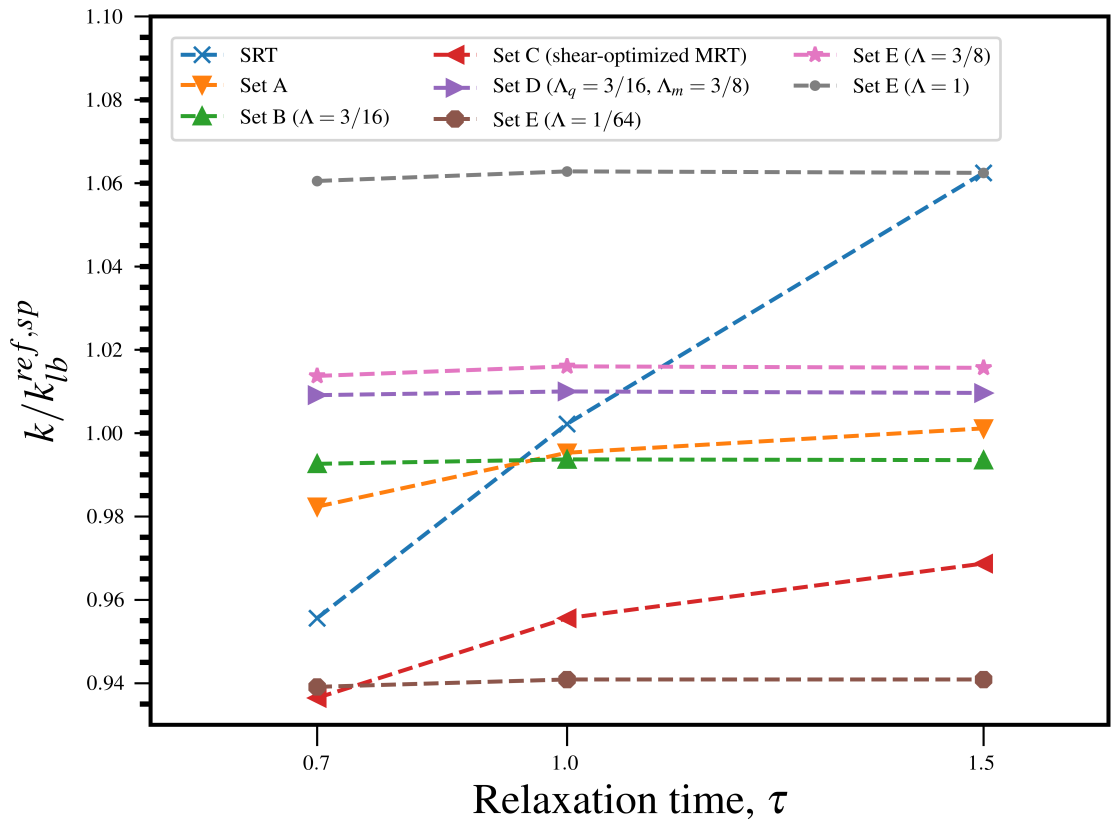}
\par\end{centering}
\caption{Normalized permeability of random packing of spheres obtained with
different sets of relaxation parameters where $k_{ref},lb=5.6426\,l.u$
or $280,151\,mD.$\label{fig:PermeabilitySpheres}}
\end{figure}

Fig. \ref{fig:PermeabilitySpheres} shows the computed normalized
permeability with various sets of parameters as a function of relaxation
time (viscosity). As observed previously, the SRT permeability predictions
increase with $\tau$ which is a a violation of Stokes flow. Predictions
with Set A and shear-optimized parameter set (Set C) vary with viscosity
but the variation is relatively small due to the (relatively) high-resolution
nature of the voxels. Set B, D, and E, on the other hand, show very
small variation in permeability with respect to viscosity which indicates
that these parameter sets can generate viscosity-independent Stokes
solutions and permeability for complex pore-spaces. We can also observe
that Set A and Set B $(\Lambda=3/16)$ result in the most accurate
predictions, followed by Set D $(\Lambda_{m}=3/16,\Lambda_{m}=3/8)$,
and Set E with $\Lambda=3/8.$ We also observe that very small $(\Lambda=1/64)$
and very large ($\Lambda=1)$ values of $\Lambda$ lead to considerably
inaccurate solutions, again, indicating that permeability predictions
by MRT collision model (and the bounceback scheme) depend on $\Lambda$.
This observation confirms the conclusions in Ref. \cite{khirevich2015coarse}
that $\Lambda\in[1/8,\,3/8]$ provides an optimal range for all resolutions
and geometries. Finally, the above results, also indicate that for
a given voxel resolution and pore-geometry, the optimal value of$\Lambda,$
in terms of accuracy, may have to be empirically determined within
the bounds $\Lambda\in[1/8,\,3/8]$. Regarding convergence rates,
similar to previous cases, we observed that increasing $\tau$ leads
to faster convergence. For instance with the Set D parameters, the
number of iterations required to converge decreased consistently from
6800 for $\tau=0.7$ to 3600 for $\tau=1.5.$ Similar trends were
observed for the other sets (A-D) of parameters.

Note that at $\tau=1,$ the SRT predictions are very close to the
reference solutions and solutions obtained from Sets B, D, and E.
This indicates that the popular option of setting $\tau=1$ with SRT
model might be produce reasonably accurate results if the voxel resolution
of the micro-tomographic images is sufficiently high. The option of
fixing $\tau=1$ for the SRT model, however, limits the option to
expedite convergence, since increasing $\tau$ to improve convergence
will quickly lead to erroneous predictions.

\section{Conclusions\label{sec:Conclusions}}

In this paper, we performed a systematic numerical evaluation of the
different sets of relaxation parameters in the MRT model for modeling
Stokes flow in 3-D microtomographic pore-spaces using the bounceback
scheme. These sets of parameters are evaluated from the point of view
of accuracy and an ability to generate viscosity-independent permeability
solutions. Instead of tuning all six independent relaxation rates
that are available in the D3Q19-MRT model, the sets that were analyzed
have relaxation rates that depend on one or two independent parameters,
namely $\tau$ and $\Lambda$. We tested elementary porous media at
different image resolutions and a random packing of spheres at relatively
high resolution. We can then conclude with the following remarks:
\begin{itemize}
\item For low-resolution micro-tomographic images, all collision models,
including the MRT model and irrespective of the relaxation rates,
produce solutions with very large numerical errors. Per Poiseuille
law, since the overall flow rate is proportional to square of the
cross-sectional area, the smallest pores have little impact of the
overall flowrate. Therefore, as long as the larger pores are sufficiently
resolved during imaging, the MRT model can produce reasonably accurate
predictions. For low-permeability media, however, unresolved smallest
pores are expected to be important.
\item For a square channel permeated with a circular bore, among the various
$\Lambda=\{\frac{1}{64},\frac{3}{16},\frac{3}{8}\}$, we observe that
$\Lambda=3/8$ results in the most accurate solutions at all resolutions
that were tested. Similarly, for a channel with triangular cross-section
we find the following $\Lambda=3/16$ gives the most accurate results.
On the other hand, for a dense random packing of identical spheres
at 7$\mu m$ resolution, we observe that $\Lambda\in[1/8,\,3/8]$.
This observation is consistent with the observations in \cite{khirevich2015coarse}.
If the very high resolution images are available, SRT with $\tau=1$
can also provide permeability with approximately the same accuracy
as MRT models. However, since it leads viscosity-dependent permeability,
the accuracy will vary depending choice of $\tau$.
\item For the MRT model, sub-optimal choices of relaxation rates can lead
to slower convergence and viscosity-dependent permeability predictions.
In fact, it was observed that the popular choice of MRT parameters,
as proposed in the seminal paper \cite{d2002multiple}, result in
permeability that depends (although weakly) on $\tau\,(\nu),$ a nonphysical
artifact for simulating Stokes flow. Among the set of parameters that
we tested, Sets B, D, and E consistently produce viscosity-independent
(parameter-independent) results for all cases, for all resolutions,
and over the entire range of $\tau$. We also observed that a larger
$\tau$ leads to a substantial reduction in the number of iterations
required for convergence without any adverse effects on accuracy.
Therefore, if a MRT implementation/code is to be used for pore-scale
Stokes flow, then it is recommended that the relaxation rates $(s_{0}-s_{18})$
should be chosen as per Sets B, D, or E (listed in Table \ref{tab:Different-sets-of})
and $\tau$ be chosen in the range $\tau\in[1.2,1.5]$, in order to
obtain viscosity-independent results, including permeability. For
Set E specifically, choosing $\Lambda\in[1/8,\,3/8]$ and $\tau\in[1.0,1.3]$
can result in overall superior accuracy, convergence rate, and parameter-independent
predictions.
\end{itemize}

\paragraph{Acknowledgments}

Financial support from the Energy and Environment Institute (EEi)
at Rice University is gratefully acknowledged. This work was supported
in part by the Data Analysis and Visualization Cyberinfrastructure
funded by the National Science Foundation (NSF) under grant OCI-0959097
and Rice University. We also thank the developers of Palabos for their
technical assistance.

\bibliographystyle{plain}
\bibliography{References}

\begin{thebibliography}{10}

\bibitem{ahrenholz2006lattice}
Benjamin Ahrenholz, Jonas Tolke, and Manfred Krafczyk.
\newblock Lattice-boltzmann simulations in reconstructed parametrized porous
  media.
\newblock {\em International Journal of Computational Fluid Dynamics},
  20(6):369--377, 2006.

\bibitem{berg2017industrial}
Carl~Fredrik Berg, Olivier Lopez, and H{\aa}vard Berland.
\newblock Industrial applications of digital rock technology.
\newblock {\em Journal of Petroleum Science and Engineering}, 157:131--147,
  2017.

\bibitem{bouzidi2001momentum}
Mohamed Bouzidi, Mouaouia Firdaouss, and Pierre Lallemand.
\newblock Momentum transfer of a boltzmann-lattice fluid with boundaries.
\newblock {\em Physics of fluids}, 13(11):3452--3459, 2001.

\bibitem{bultreys2016imaging}
Tom Bultreys, Wesley De~Boever, and Veerle Cnudde.
\newblock Imaging and image-based fluid transport modeling at the pore scale in
  geological materials: A practical introduction to the current
  state-of-the-art.
\newblock {\em Earth-Science Reviews}, 155:93--128, 2016.

\bibitem{d2002multiple}
Dominique dHumi{\`e}res.
\newblock Multiple--relaxation--time lattice boltzmann models in three
  dimensions.
\newblock {\em Philosophical Transactions of the Royal Society of London A:
  Mathematical, Physical and Engineering Sciences}, 360(1792):437--451, 2002.

\bibitem{d2009viscosity}
Dominique dHumieres and Irina Ginzburg.
\newblock Viscosity independent numerical errors for lattice boltzmann models:
  from recurrence equations to magic collision numbers.
\newblock {\em Computers \& Mathematics with Applications}, 58(5):823--840,
  2009.

\bibitem{eshghinejadfard2016calculation}
Amir Eshghinejadfard, L{\'a}szl{\'o} Dar{\'o}czy, G{\'a}bor Janiga, and
  Dominique Th{\'e}venin.
\newblock Calculation of the permeability in porous media using the lattice
  boltzmann method.
\newblock {\em International Journal of Heat and Fluid Flow}, 62:93--103, 2016.

\bibitem{fattahi2016lattice}
Ehsan Fattahi, Christian Waluga, Barbara Wohlmuth, Ulrich R{\"u}de, Michael
  Manhart, and Rainer Helmig.
\newblock Lattice boltzmann methods in porous media simulations: From laminar
  to turbulent flow.
\newblock {\em Computers \& Fluids}, 140:247--259, 2016.

\bibitem{Ferrol1995}
Bruno Ferr{\'e}ol and Daniel~H. Rothman.
\newblock Lattice-boltzmann simulations of flow through fontainebleau
  sandstone.
\newblock {\em Transport in Porous Media}, 20(1):3--20, Aug 1995.

\bibitem{Finney}
John Finney.
\newblock Finney packing of spheres.
\newblock \url{http://www.digitalrocksportal.org/projects/47}, 2016.

\bibitem{fredrich2006predicting}
JT~Fredrich, AA~DiGiovanni, and DR~Noble.
\newblock Predicting macroscopic transport properties using microscopic image
  data.
\newblock {\em Journal of Geophysical Research: Solid Earth}, 111(B3), 2006.

\bibitem{ginzbourg1994boundary}
I~Ginzbourg and PM~Adler.
\newblock Boundary flow condition analysis for the three-dimensional lattice
  boltzmann model.
\newblock {\em Journal de Physique II}, 4(2):191--214, 1994.

\bibitem{ginzburg2006variably}
Irina Ginzburg.
\newblock Variably saturated flow described with the anisotropic lattice
  boltzmann methods.
\newblock {\em Computers \& fluids}, 35(8-9):831--848, 2006.

\bibitem{ginzburg2003multireflection}
Irina Ginzburg and Dominique dHumieres.
\newblock Multireflection boundary conditions for lattice boltzmann models.
\newblock {\em Physical Review E}, 68(6):066614, 2003.

\bibitem{ginzburg2008two}
Irina Ginzburg, Frederik Verhaeghe, and Dominique dHumieres.
\newblock Two-relaxation-time lattice boltzmann scheme: About parametrization,
  velocity, pressure and mixed boundary conditions.
\newblock {\em Communications in computational physics}, 3(2):427--478, 2008.

\bibitem{guo2007discrete}
Zhaoli Guo, Baochang Shi, TS~Zhao, and Chuguang Zheng.
\newblock Discrete effects on boundary conditions for the lattice boltzmann
  equation in simulating microscale gas flows.
\newblock {\em Physical Review E}, 76(5):056704, 2007.

\bibitem{guo2013lattice}
Zhaoli Guo and Chang Shu.
\newblock {\em Lattice Boltzmann method and its applications in engineering},
  volume~3.
\newblock World Scientific, 2013.

\bibitem{khirevich2015coarse}
Siarhei Khirevich, Irina Ginzburg, and Ulrich Tallarek.
\newblock Coarse-and fine-grid numerical behavior of mrt/trt lattice-boltzmann
  schemes in regular and random sphere packings.
\newblock {\em Journal of Computational Physics}, 281:708--742, 2015.

\bibitem{kruger2017lattice}
Timm Kr{\"u}ger, Halim Kusumaatmaja, Alexandr Kuzmin, Orest Shardt, Goncalo
  Silva, and Erlend~Magnus Viggen.
\newblock The lattice boltzmann method, 2017.

\bibitem{lallemand2000theory}
Pierre Lallemand and Li-Shi Luo.
\newblock Theory of the lattice boltzmann method: Dispersion, dissipation,
  isotropy, galilean invariance, and stability.
\newblock {\em Physical Review E}, 61(6):6546, 2000.

\bibitem{luo2010lattice}
Li-Shi Luo, Manfred Krafczyk, and Wei Shyy.
\newblock Lattice boltzmann method for computational fluid dynamics.
\newblock {\em Encyclopedia of Aerospace Engineering}, pages 651--660, 2010.

\bibitem{luo2011numerics}
Li-Shi Luo, Wei Liao, Xingwang Chen, Yan Peng, Wei Zhang, et~al.
\newblock Numerics of the lattice boltzmann method: Effects of collision models
  on the lattice boltzmann simulations.
\newblock {\em Physical Review E}, 83(5):056710, 2011.

\bibitem{maier2010lattice}
RS~Maier and RS~Bernard.
\newblock Lattice-boltzmann accuracy in pore-scale flow simulation.
\newblock {\em Journal of Computational Physics}, 229(2):233--255, 2010.

\bibitem{manwart2002lattice}
C~Manwart, U~Aaltosalmi, A~Koponen, R~Hilfer, and J~Timonen.
\newblock Lattice-boltzmann and finite-difference simulations for the
  permeability for three-dimensional porous media.
\newblock {\em Physical Review E}, 66(1):016702, 2002.

\bibitem{martys1996simulation}
Nicos~S Martys and Hudong Chen.
\newblock Simulation of multicomponent fluids in complex three-dimensional
  geometries by the lattice boltzmann method.
\newblock {\em Physical Review E}, 53(1):743, 1996.

\bibitem{mattila2016prospect}
Keijo Mattila, Tuomas Puurtinen, Jari Hyv{\"a}luoma, Rodrigo Surmas, Markko
  Myllys, Tuomas Turpeinen, Fredrik Robertsen, Jan Westerholm, and Jussi
  Timonen.
\newblock A prospect for computing in porous materials research: Very large
  fluid flow simulations.
\newblock {\em Journal of Computational Science}, 12:62--76, 2016.

\bibitem{mavko2009rock}
Gary Mavko, Tapan Mukerji, and Jack Dvorkin.
\newblock {\em The rock physics handbook: Tools for seismic analysis of porous
  media}.
\newblock Cambridge university press, 2009.

\bibitem{narvaez2010quantitative}
Ariel Narv{\'a}ez, Thomas Zauner, Frank Raischel, Rudolf Hilfer, and Jens
  Harting.
\newblock Quantitative analysis of numerical estimates for the permeability of
  porous media from lattice-boltzmann simulations.
\newblock {\em Journal of Statistical Mechanics: Theory and Experiment},
  2010(11):P11026, 2010.

\bibitem{pan2006evaluation}
Chongxun Pan, Li-Shi Luo, and Cass~T Miller.
\newblock An evaluation of lattice boltzmann schemes for porous medium flow
  simulation.
\newblock {\em Computers \& fluids}, 35(8-9):898--909, 2006.

\bibitem{BenchmarkDatabase}
Nishank Saxena.
\newblock Data for: References and benchmarks for pore-scale flow simulated
  using micro-ct images of porous media and digital rocks, 2017.

\bibitem{saxena2017references}
Nishank Saxena, Ronny Hofmann, Faruk~O Alpak, Steffen Berg, Jesse Dietderich,
  Umang Agarwal, Kunj Tandon, Sander Hunter, Justin Freeman, and Ove~Bjorn
  Wilson.
\newblock References and benchmarks for pore-scale flow simulated using
  micro-ct images of porous media and digital rocks.
\newblock {\em Advances in Water Resources}, 109:211--235, 2017.

\bibitem{talon2012assessment}
L~Talon, D~Bauer, N~Gland, S~Youssef, H~Auradou, and I~Ginzburg.
\newblock Assessment of the two relaxation time lattice-boltzmann scheme to
  simulate stokes flow in porous media.
\newblock {\em Water Resources Research}, 48(4), 2012.

\bibitem{Palabos}
Palabos Team.
\newblock Palabos:open-source lbm software.
\newblock \url{http://www.palabos.org/}, 2009--2017.

\bibitem{wildenschild2013x}
Dorthe Wildenschild and Adrian~P Sheppard.
\newblock X-ray imaging and analysis techniques for quantifying pore-scale
  structure and processes in subsurface porous medium systems.
\newblock {\em Advances in Water Resources}, 51:217--246, 2013.

\bibitem{xu2016novel}
Lina Xu, Parthib Rao, and Laura Schaefer.
\newblock A novel scheme for curved moving boundaries in the lattice boltzmann
  method.
\newblock {\em International Journal of Modern Physics C}, 27(12):1650144,
  2016.

\bibitem{young2008efficient}
PG~Young, TBH Beresford-West, SRL Coward, B~Notarberardino, B~Walker, and
  A~Abdul-Aziz.
\newblock An efficient approach to converting three-dimensional image data into
  highly accurate computational models.
\newblock {\em Philosophical Transactions of the Royal Society of London A:
  Mathematical, Physical and Engineering Sciences}, 366(1878):3155--3173, 2008.

\bibitem{ZARETSKIY20101508}
Yan Zaretskiy, Sebastian Geiger, Ken Sorbie, and Malte Forster.
\newblock Efficient flow and transport simulations in reconstructed 3d pore
  geometries.
\newblock {\em Advances in Water Resources}, 33(12):1508 -- 1516, 2010.

\end{thebibliography}

\end{document}